\documentclass[twocolumn,aps,prb]{revtex4}
\usepackage{amsmath,empheq}
\usepackage{amssymb}
\usepackage{mathrsfs}  
\usepackage{amsfonts}
\usepackage{booktabs}
\usepackage{physics}
\usepackage{graphicx} 
\usepackage{subfigure} 
\usepackage{epsfig}
\usepackage{color}
\usepackage{epstopdf}
 \usepackage{rotating}
 \usepackage{appendix}
 \usepackage{lscape}
 \usepackage[colorlinks=true,linkcolor=blue,citecolor=blue,filecolor=blue,urlcolor=blue]{hyperref}

\begin{document}

 \title{Magnetism and topological phases in an interacting decorated honeycomb lattice with spin-orbit coupling}
 \author{Manuel Fern\'andez L\'opez and Jaime Merino}
\affiliation{Departamento de F\'isica Te\'orica de la Materia Condensada, Condensed Matter Physics Center (IFIMAC) and
Instituto Nicol\'as Cabrera, Universidad Aut\'onoma de Madrid, Madrid 28049, Spain}
\begin{abstract}
We study the interplay between spin-orbit coupling (SOC) and Coulomb repulsion in a Hubbard model on a decorated honeycomb lattice which leads to a plethora of phases. While a quantum spin hall insulator is stable 
at weak Coulomb repulsion and moderate SOC, a semimetallic phase emerges at large SOC in a broad range of Coulomb repulsion.
This semimetallic phase has topological properties not observed in conventional metals such as a finite, non-quantized spin Hall conductivity. At large Coulomb repulsion and negligible spin-orbit coupling, electronic correlations stabilize a resonance valence bond (RVB) spin liquid state in contrast to the classical antiferromagnetic state predicted by mean-field theory. Under sufficiently strong SOC, such RVB state is 
transformed into a magnetic insulator consisting on $S \lesssim 3/2$ localized moments on a honeycomb lattice with antiferromagnetic order and topological features.
\end{abstract}
 \date{\today}

\maketitle
 
\section{Introduction}

Since the discovery of topological insulators 
there is an intense research 
activity around spin-orbit coupling (SOC) effects on materials.\cite{Discovered} 
The quantum spin hall (QSH) phase\cite{KaneMele} arising in weakly interacting electron systems is well understood by now, however, much less is known about possible new phases arising from the interplay between SOC and Coulomb repulsion in strongly correlated materials\cite{balents}. This is relevant to Ir-based pyrochlores in which a topological Mott insulator (TMI), axion insulator, topological Weyl semi-metal and quantum spin liquids\cite{pesin,kim} can occur. Many of these phases are characterized by having topological order {\it i .e.} long range entanglement\cite{wen}, rather than being protected by topological invariants as in conventional topological insulators. 

The transition from a topological insulator to a Mott insulator has been explored in the Kane-Mele-Hubbard (KMH)
model\cite{lehur,hohenadler}. With no SOC, a transition from a SM  to a N\'eel ordered Mott insulator through a quantum spin liquid (QSL) occurs\cite{assaad}. For any non-zero SOC, the non-interacting QSH insulator with metallic edges is stable up to weak Coulomb repulsion. As Coulomb repulsion is further increased up to intermediate values, a transition from the QSH to a Mott insulator with easy-axis AF occurs. The QSH phase is stabilized in a broader range of Coulomb repulsion with the increase of SOC. 

The decorated honeycomb lattice is relevant to many materials such as trinuclear organometallic compounds\cite{khosla,jacko,merino,powell} {\it e. g.} Mo$_3$S$_7$(dmit)$_3$, organic molecular crystals, Iron (III) acetates\cite{iron3}, molecular organic frameworks \cite{henline,henling} (MOFs) and cold fermionic atoms loaded in optical lattices\cite{cdmft}. One may expect interesting physical phenomena arising from the frustration of the lattice which interpolates between the honeycomb and Kagom\'e lattices\cite{jacko1}. Indeed, a non-interacting tight-binding model with SOC, topological phases such as the QSH arise\cite{ruegg}. In a spinless extended Hubbard model, the off-site Coulomb repulsion leads to a QAH insulator and/or to a topological metal which breaks TRS spontaneously even when no SOC is present\cite{int,scarola,manuel}. Finally, the Hubbard model on the DHL hosts a broad variety of phases including: a real space Mott insulator at half-filling, trimer and dimer Mott insulator as well as a spin triplet Mott at $4/3$-filling.\cite{Nourse}  In the strong coupling limit, when local spin-$1/2$ moments have formed, VBS \cite{Heisenberg,Orus} and quantum spin liquids\cite{kivelson} can be stabilized. Hence, a plethora of phases arise in a Hubbard model with a single-orbital per site typically associated with multi-orbital Hubbard models.

Here, we take a step beyond previous work on the DHL by considering interacting and spinful electrons. In order to do this, we explore the phase diagram of a Hubbard model on the DHL with SOC based on mean-field Hartree-Fock (HF) and exact diagonalization (ED) techniques. The non-interacting QSH and SM phases\cite{ruegg} are found to be stable up to SOC dependent $U_c(\lambda_{SO})$ that increases with $\lambda_{SO}$ similar to the behavior in the KMH model but in contrast to a model for Ir-pyrochlores.\cite{pesin,kim} At weak SOC, a transition from a QSH to a quantum spin liquid (QSL) phase induced by $U$ occurs. In contrast, HF would find a transition to a classical AFI phase. At strong SOC, the SM phase, which is stable in a broad weak-to-intermediate $U$ repulsion regime, is characterized by a finite non-quantized spin Hall conductivity (in contrast to the quantization associated with the QSH phase). At sufficiently strong $U$, a transition from the SM to a magnetically ordered phase with localized $S \lesssim 3/2$ moments at each triangle occurs. At the mean-field level, this phase effectively is a $S=3/2$ AF on the honeycomb lattice. The phase diagram obtained is richer than the one of the KMH model including new phases (SM and $3/2$-MI) and phase transitions.

Our work is organized as follows. In Sec. \ref{sec:methods} the model and the Hartree-Fock method are introduced. In Sec. \ref{sec:pd} the $U-\lambda_{SO}$ mean field phase diagram at half-filling is  obtained in Sec. \ref{sec:pd}. In Sec. \ref{sec:ed} electron correlation effects on the mean-field states are explored using ED techniques. In Sec. \ref{sec:exp} we provide an analysis of the spin hall conductivity which could be compared with experiments. Finally, in Sec. \ref{sec:concl} we conclude the paper giving an outlook of future works. 
\section{Model and methods}
\label{sec:methods}
We consider a Hubbard model under the effect of SOC on the decorated honeycomb lattice:
\begin{align}
\begin{aligned}
&\mathcal{H} = \mathcal{H}_{tb}+\mathcal{H}_{Coul}+\mathcal{H}_{SO}, \\
&\mathcal{H}_{tb} = -t \sum_{\langle ij\rangle,\sigma }c_{i\sigma}^{\dagger}c_{j\sigma}, \\
&\mathcal{H}_{Coul} = U\sum_{i}n_{i\uparrow}n_{i\downarrow},\\
&\mathcal{H}_{SO} = {i}\lambda_{SO}\sum_{\langle\langle ij \rangle\rangle} e_{ij} (c^\dagger_{i\uparrow}c_{j\uparrow} - c^\dagger_{i\downarrow} c_{j\downarrow}), 
\end{aligned}
\label{Hamiltonian}
\end{align}
where $c_{i\sigma}^\dagger$ ($c_{i\sigma}$) creates (annihilates) a fermion on site $i$ with spin $\sigma=\uparrow,\downarrow$. The hopping amplitude $t$ is the same for n.n. inside and between triangles in the lattice while the n.n.n. hopping amplitude induced by SOC $i\lambda_{SO}$ changes sign for right (left) turning electrons as encoded in $e_{ij}= +1 (-1)$ \cite{KaneMele}. In the present work, we will be interested in the different phases arising in the model for different $U$ and $\lambda_{SO}$ at half-filling.

\subsection{Hartree-Fock approach}

When we switch the on-site Coulomb repulsion, $U\neq 0$, the hamiltonian becomes cuartic. Since it cannot be solved exactly we apply a Hartree-Fock mean-field decoupling of the Coulomb interaction:
\begin{align}
    n_{i\uparrow}n_{i\downarrow}&\approx (n_{i\uparrow}n_{i\downarrow})_{Hartree}-(n_{i\uparrow}n_{i\downarrow})_{Fock} 
\end{align} 
where $(n_{i\uparrow}n_{i\downarrow})_{Hartree}=n_{i\uparrow} \langle n_{i\downarrow}\rangle +  \langle n_{i\uparrow}\rangle n_{i\downarrow}-\langle n_{i\uparrow} \rangle\langle n_{i\downarrow}\rangle$ and $(n_{i\uparrow}n_{i\downarrow})_{Fock}=c_{i\uparrow}^\dagger c_{i\downarrow}\langle c_{i\downarrow}^\dagger c_{i\uparrow}\rangle + \langle c_{i\uparrow}^\dagger c_{i\downarrow}\rangle c_{i\downarrow}^\dagger c_{i\uparrow} - \langle c_{i\uparrow}^\dagger c_{i\downarrow}\rangle \langle c_{i\downarrow}^\dagger c_{i\uparrow}\rangle$. We work in the canonical ensemble with a fixed number of electrons $N_e$. At a given temperature $\frac{1}{\beta}=k_B T$, the free energy $\mathcal{F}$ is given by $\mathcal{F}=\mathcal{F}_T-\mathcal{F}_{HF}$, where:
\begin{align}
\label{FreeT}
&\mathcal{F}_T=-k_BT\sum_{\textbf{k},n}log[1+e^{-\beta(E_{\textbf{k},n}-\mu)}]+\mu N_e&  \\
\label{FreeH}
&\mathcal{F}_{HF}=U\sum_i \left(\langle n_{i\uparrow}\rangle \langle n_{i\downarrow}\rangle - \langle c_{i\uparrow}^\dagger c_{i\downarrow}\rangle \langle c_{i\downarrow}^\dagger c_{i\uparrow}\rangle \right)&
\end{align}
with $\mu$ the chemical potential and $E_{\textbf{k},n}$ the $n^{th}$ Hartree-Fock energy band. We consider complex Fock terms $\langle c_{i\uparrow}^\dagger c_{i\downarrow}\rangle = \xi_i+i\eta_i$ in such a way that $\langle c_{i\uparrow}^\dagger c_{i\downarrow}\rangle \langle c_{i\downarrow}^\dagger c_{i\uparrow}\rangle=\xi_i^2+\eta_i^2$. These terms correspond to the spin $x$ and $y$ components respectively $\eta_i = S_i^y$ and $\xi_i=S_i^x$ since $c_{i\uparrow}^\dagger c_{i\downarrow}^\dagger=S^+_i$ and $S^+_i=S_i^x+iS_i^y$. Carrying out the minimization of $\mathcal{F}$ with respect each Hartree-Fock variable, we get a set of 24 coupled self-consistent equations:
\begin{align}
\left.\begin{aligned}
\langle n_{i\uparrow}\rangle^{(w+1)}=\frac{1}{U}\sum_{{\bf k}, n}\frac{\partial E_{{\bf k}, n}/\partial n_{i\downarrow}^{(w)}}{1+e^{\beta (E_{{\bf k}, n}-\mu)}}\\
\langle n_{i\downarrow}\rangle^{(w+1)}=\frac{1}{U}\sum_{{\bf k}, n}\frac{\partial E_{{\bf k}, n}/\partial n_{i\uparrow}^{(w)}}{1+e^{\beta (E_{{\bf k}, n}-\mu)}}\\
\xi_{i}^{(w+1)}=-\frac{1}{2U}\sum_{{\bf k}, n}\frac{\partial E_{{\bf k}, n}/\partial \xi_{i}^{(w)}}{1+e^{\beta (E_{{\bf k}, n}-\mu)}}\\
\eta_{i}^{(w+1)}=-\frac{1}{2U}\sum_{{\bf k}, n}\frac{\partial E_{{\bf k}, n}/\partial \eta_{i}^{(w)}}{1+e^{\beta (E_{{\bf k}, n}-\mu)}}
\end{aligned}\right.
\label{selfconsistent}
\end{align}
where $1\leq i\leq N_s=6$ and $w$ is the iteration. By solving them simultaneously for each set of parameters $(\lambda_{SO},U)$, we are able to find the ground state of the system. 

In addition, we calculate the correlations matrix $\expval{c_{i\alpha}^\dagger c_{j\beta}}{\Psi_{HF}}$. The imaginary part of the non-diagonal terms will give us the elemental current between two sites of the unit cell. Then we can see how the chiral (or not-chiral) currents seem in the different phases, obtained additional information. First we write the Fourier transform of $\langle c_{i\alpha}^\dagger c_{j\beta} \rangle$ taking into account the translational invariance:
\begin{equation}
\langle c_{i \alpha}^\dagger c_{j\beta} \rangle=\frac{1}{N_s}\sum_{{\bf k}}e^{-i{\bf k\cdot (d_i-d_j)}}\langle c^\dagger_{  i  \alpha} ( {\bf k}) c_{j \beta} ({\bf k})\rangle
\label{eqoriginal}
\end{equation}
where $i,j$ are the indices for sites and $\alpha,\beta$ for spins. Now we have to change to the basis in which the HF hamiltonian is diagonalized. In this basis the creation (anihilation) operators, $b^\dagger_{n \gamma}({\bf k})$ ($b_{m \delta}({\bf k})$), are related with the ordinary fermionic operator as follows:
\begin{align}
\left.
\begin{aligned}
b^\dagger_{n \gamma}({\bf k})=\sum_{i \alpha }^{N}\xi_{n\gamma}^{i \alpha}({\bf k})c^\dagger_{i\alpha}({\bf k}) \\
b_{m \delta}({\bf k})=\sum_{j \beta}^{N}{\xi_{m\delta}^{ j \beta }({\bf k})}^*c_{m \beta}({\bf k})    
\end{aligned}
\right.
\end{align}
where the spin and bands are labeled by $\gamma ,\delta $ and $n,m$ respectively, and $\xi({\bf k}$) is the eigenvectors matrix obtained from the self-consistency. In other words, what we have done is a basis rotation in which the new operator can be written as a linear combination of the original ones. Being the global system eigenvector $\ket{\Psi_{HF}}=\prod_{n \gamma}^{N_b}\prod_{\bf k}b^\dagger_{n \gamma}({\bf k})\ket{0}$ and writing $\langle c^\dagger_{  n  \alpha} ( {\bf k}) c_{m \beta} ({\bf k})\rangle$ in terms of $b_{n'\gamma}({\bf k})^\dagger$ ($b_{m'\delta}(\bf k)$) operators
we obtain the following expression for the correlation matrix:
\begin{equation}
\langle c_{i \alpha}^\dagger c_{j\beta} \rangle=\frac{1}{N_s}\sum_{\bf k}\sum_{n \delta}e^{-i{\bf k\cdot (d_i-d_j)}}\chi_{i\alpha}^{n\delta}({\bf k})\chi_{j\beta}^{n\delta}({\bf k})^*
\label{correlation}
\end{equation}
where $\chi({\bf  k})\equiv\xi^{-1}({\bf  k})$. The elements in the diagonal are strictly real and represent the on-site $i$ electronic mean density $\langle n_{i\uparrow}\rangle (\langle n_{i\downarrow}\rangle)$. In the non-diagonal terms (which can be complex), we distinguish the on-site correlations $\langle c^\dagger_{i\uparrow} c_{i\downarrow}\rangle$, which match with $\xi_i+i\eta_i$  found by the self-consistency \eqref{selfconsistent}, and the off-site correlations $\langle c^\dagger_{i\alpha} c_{j\beta}\rangle$ with $i\neq j$. The off-site correlations can be either exchange terms (if the spin changes from one site to the other) or ordinary terms (if not). The real part acts as a hopping shift between the involved sites while the imaginary part represents the elemental current.

\subsection{Topological properties}
It is important to distinguish between a conventional and a topological insulator with a QSH phase. A gap opening is a necessary but not sufficient condition for having a topological insulator. In order to ensure this, we calculate the Z$_2$ bulk invariant using a method developed for systems with inversion spatial symmetry\cite{Z2}. This procedure consists in obtaining $Z_2$ through the parity eigenvalues of the occupied bands evaluated at the TRIM, $\Gamma_j$ points. In two dimensional systems there are four of these points:
$\overrightarrow{\Gamma}=(0,0)$, $\overrightarrow{M_1}=\frac{\overrightarrow{b_1}}{2}=\frac{\pi}{3l}(1,\sqrt{3})$, $\overrightarrow{M_2}=\frac{\overrightarrow{b_2}}{2}=\frac{\pi}{3l}(1,-\sqrt{3})$ and $\overrightarrow{M_3}=\frac{\overrightarrow{b_1}+\overrightarrow{b_2}}{2}=\frac{2\pi}{3l}(1,0)$. More specifically, we have to calculate the product of the parity eigenvalues at each $\Gamma$ point ($\delta_j$) for all occupied bands:
\begin{equation}
\delta_j=\prod_{m=1}^{N_c}\xi_{2m}\left(\Gamma_j\right)
\label{delta}
\end{equation}
where $N_c$ is the number of occupation bands and $\xi$ is just the eigenvalue of the parity operator which takes $-1$ ($+1$) if the eigenstate changes sign (or not) when the parity transformation, $\overrightarrow{r}$ by $-\overrightarrow{r}$, takes place. Hence, we can label the states by $2m$ when IS is preserved in the model which is not the case for large $U$ (AFI or $3/2$-MI) as we will see. It is neither possible to compute it when the system is immersed in a gapless state (SM). From $\delta_j$ one can compute the $Z_2$ invariant ($\nu$):
\begin{equation}
(-1)^\nu=\prod_j\delta_j
\label{nu}
\end{equation} 
On this way if $\nu=0$ the system will be just a band insulator while if $\nu=1$ it will have non-trivial topology finding itself in the QSH phase. The $Z_2$ calculation is not the only way to search for topological signatures, one can also compute the spin Chern numbers $c_{\sigma n}$\cite{Lieb}. The procedure consists in projecting the wavevectors on each spin subspace ($P_\pm =\frac{1\pm\sigma_z}{2}$) and carrying out the same numerical procedure than in the spinless case \cite{Fukuki, manuel}. On this way we are not just able to characterize topological features inside the QSH phase but also in the SM region by getting the spin Berry phases $\gamma_{\sigma n}$ associated with each Fermi surface. The spin Hall conductivity $\sigma_{xy}^s$ can be computed from these quantities \eqref{condformula} providing experimental evidences of these topological  states as it is detailed in Sec. \ref{sec:exp}.
\section{Mean-field analysis of the Hubbard model with SOC on the decorated honeycomb lattice}
\label{sec:pd}
In this section we analyze the different ground states arising
in the Hubbard model in presence of SOC on the DHL under a
Hartree-Fock (HF) treatment. The different ground states of this model at half-filling are obtained by solving the HF equations 
\eqref{selfconsistent} for fixed parameters: $U,\lambda_{SO}$. In this way we construct the $U-\lambda_{SO}$ phase diagram of the
model at half-filling. We characterize the electronic, magnetic and topological properties of the different ground states obtained as well as the transitions occurring between them. 
\begin{figure}[!b]
   \centering
   \includegraphics[width=7.5cm]{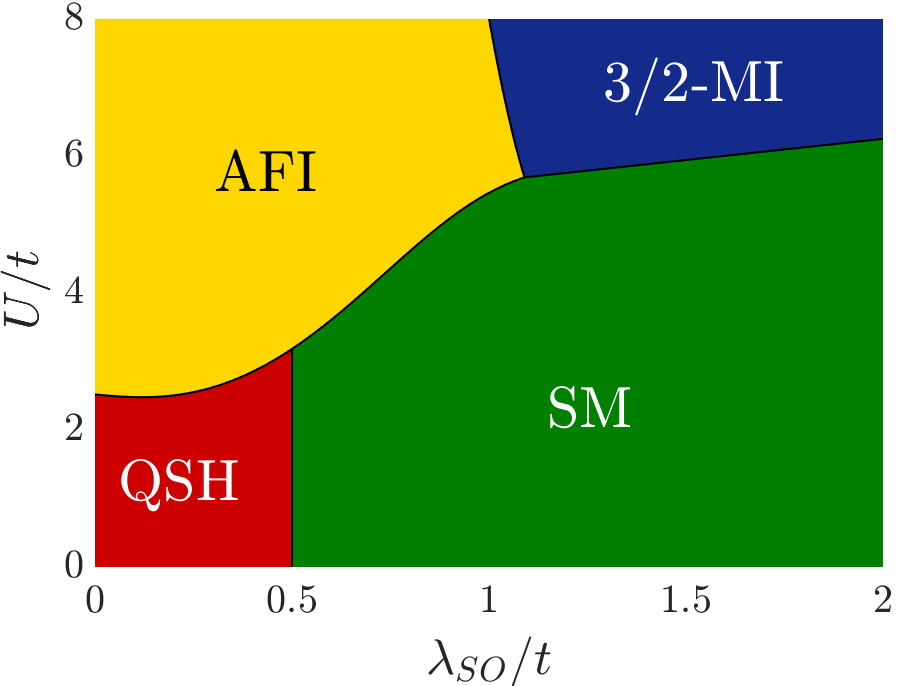}
       \caption{The $U$ vs $\lambda_{SO}$ Hartree-Fock phase diagram of the half-filled Hubbard model on the decorated honeycomb lattice with SOC. We find two non-magnetic phases: the gapped QSH and the gapless SM, and, two insulating magnetic phases: the AFI and $3/2$-MI phases.
       The QSH phase and the gapless SM phase preserve TRS and IS whereas the AFI and the $3/2$-MI phases break spontaneously both symmetries through spin ordering phenomena which patterns are shown in Fig. \ref{Schemes}.
       All four-phases contain non-zero bond currents as can be seen in Fig. \ref{Schemes} for the spin-up sector, $\langle c_{i\uparrow} c_{j\uparrow}\rangle$.}
 \label{HFSOCUdiagram}
 \end{figure} 
 \subsection{Phase diagram}
\begin{figure}[b!]
\centering
\subfigure[]{\includegraphics[width=3.5cm]{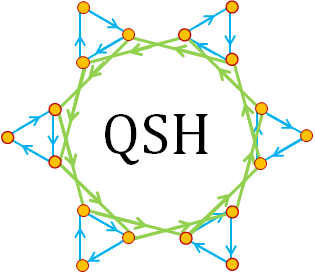}}
\hspace{0.7cm}
\subfigure[]{\includegraphics[width=3.5cm]{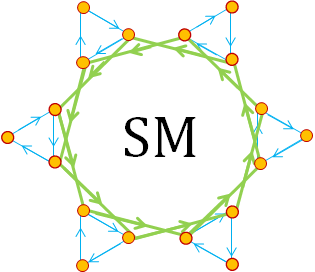}}
\subfigure[]{\includegraphics[width=3.5cm]{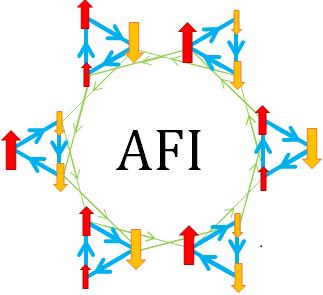}}
\hspace{0.7cm}
\subfigure[]{\includegraphics[width=3.5cm]{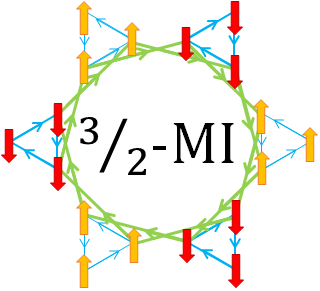}}
      \caption{Hartree-Fock ground states of the half-filled Hubbard model on the decorated honeycomb lattice with SOC. 
      (a) Quantum spin Hall (QSH) phase: topological insulator characterized by the $Z_2$ invariant, $\nu=1$, protected by TRS and IS. The nonzero chiral currents in the spin-$\uparrow$ sector shown cancel the spin-$\downarrow$ currents (not shown) preserving the TRS.
      (b) Semimetallic (SM) phase: The IS and TRS are kept but the Fermi level cross two of the bands (Fig. \ref{QSHBItransition}{\color{blue}(c)}). This phase is characterized by a non-zero spin Chern number and nonzero Berry phase associated with the chiral currents shown. A nonzero spin Hall conductivity (Fig. \ref{spinHallconductivity}) can be associated with this SM phase. (c) Antiferromagnetic insulator (AFI) phase. TRS is broken since spins are ordered following the color pattern being the inter-triangle n.n. spins always opposite.
      (d) Effective spin-$3/2$ magnetic insulator ($3/2$-MI) phase. The spins belonging to the same triangle are equally oriented and opposite to the n.n. triangle forming an effective spin, $S=3/2$ antiferromagnetic system. The third spin component is represented as red for $\uparrow$ and yellow for $\downarrow$ and the amplitude depends on the size of the arrow. The currents between sites are represented as green arrows for n.n.n. loops and blue arrows for n.n. loops and the thickness reflects the current amplitude. For phases with uniform spin distribution, this is, $\langle n_{i\uparrow}\rangle=\langle n_{i\downarrow}\rangle=0.5$ (QSH and SM), $\langle S_i^z \rangle$ are represented as yellow circles.}
\label{Schemes}
\end{figure}

The $U-\lambda_{SO}$ phase diagram of our model \eqref{Hamiltonian} is shown in Fig. \ref{HFSOCUdiagram}. The four different phases found are: a quantum spin Hall (QSH), a semimetallic (SM), an antiferromagnetic insulator (AFI) and  effective spin $S=3/2$ magnetic insulator ($3/2$-MI). The real space configurations of these states are schematically illustrated in Fig. \ref{Schemes}. They are characterized by the value of the third spin component $\langle S_i^z \rangle={\langle n_{i\uparrow}\rangle-\langle n_{i\downarrow}\rangle \over 2}$ at each lattice site $i$ and the average n.n. and n.n.n. bond currents, $\langle c_{i \uparrow}^\dagger c_{j\uparrow} \rangle$, obtained from \eqref{correlation}. While the QSH and SM phases are non-magnetic, the AFI and $3/2$-MI phases do order magnetically as shown in Fig. \ref{Schemes}. On the other hand all phases do sustain spontaneous non-zero bond currents of some kind. We now discuss each of these phases based on their magnetic properties and their electronic band structure.

\subsection{QSH phase}
We find a QSH phase in the small $U, \lambda_{SO}$ region of the phase diagram shown in Fig. \ref{HFSOCUdiagram} consistent with the QSH phase encountered previously in the non-interacting tight-binding model ($U=0$) with SOC.\cite{ruegg} The QSH is stable for sufficiently weak $U \lesssim 2.5 t$ in a broad SOC range: $0< \lambda_{SO} < 0.7 t$. {The non-interacting DHL in absence of SOC ($U, \lambda_{SO}=0$) is unstable against SOC leading to a QSH similar to one predicted in the honeycomb lattice. \cite{KaneMele}} 

\begin{figure}[b!]
   \centering
   \includegraphics[width=2.8cm]{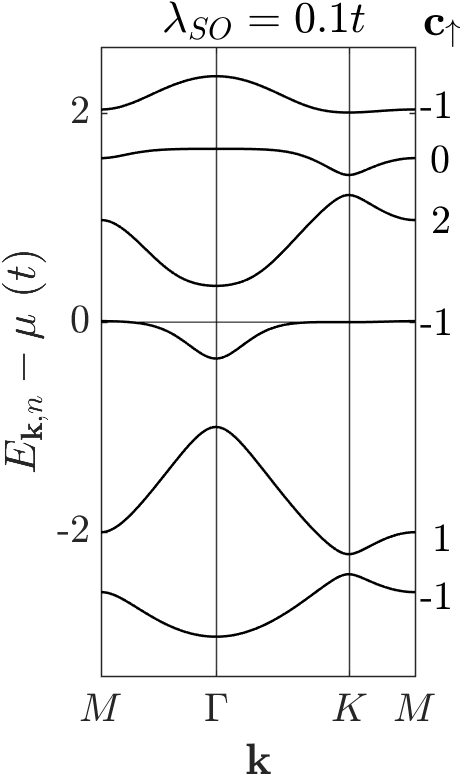}
   \includegraphics[width=2.8cm]{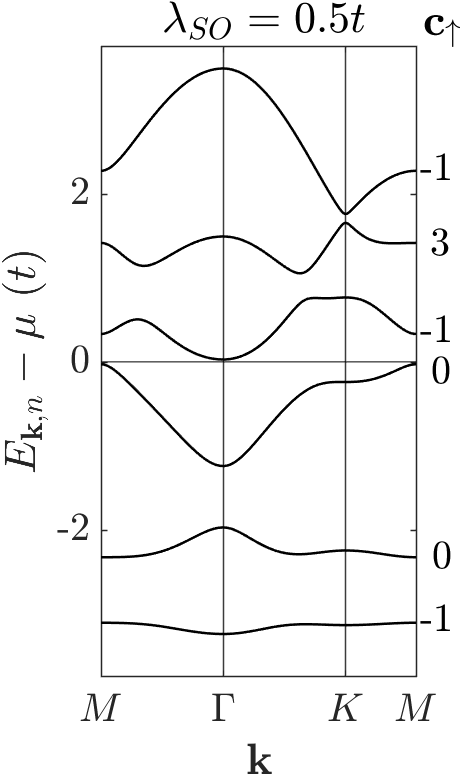} 
   \includegraphics[width=2.8cm]{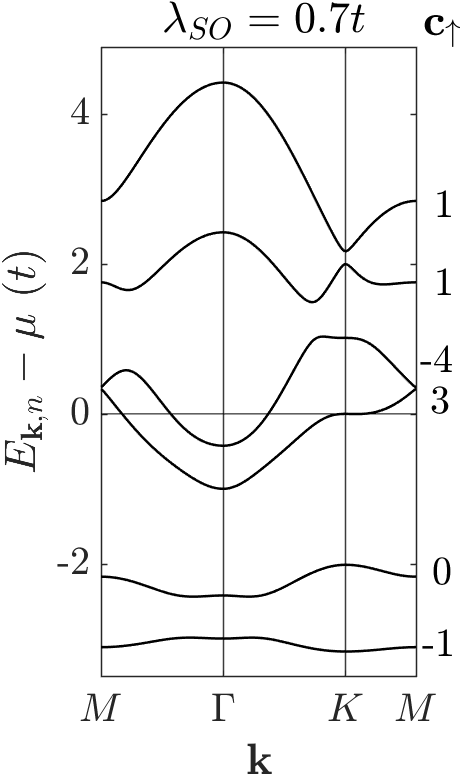}
       \caption{Hartree-Fock electron band dispersions at weak Coulomb repulsion, $U=2t$, across the QSH to SM transition. (a) $\lambda_{SO}=0.1t$. A gap is opened up at $\Gamma$ and the system turns into a topological insulator with the $Z_2$ bulk invariant, $\nu=1$, (b) $\lambda_{SO}=0.5t$. The band dispersion are strongly modified by SOC so that the fourth band almost touches the Fermi level. The spin Chern numbers,  $c_\uparrow$, from the second to fifth bands change due to a topological phase transition occurring at $\lambda_{SO}\sim 0.3t$ involving the closing of the gap between the second and third bands at the $\Gamma$-point and between fourth and fifth bands at the $K$-point. (c) $\lambda_{SOC}=0.7t$. The gapless SM having the third and fourth partially filled bands has formed. Again, the $c_\uparrow$ have been altered due to the gap closing at $K$ ($M$) between the (fifth and sixth) third and fourth bands 
       at $\lambda_{SO}\sim 0.6t$ ($\lambda_{SO}\sim 0.7t$).
       }
 \label{QSHBItransition}
 \end{figure}

This topological QSH state is characterized by dissipationless currents associated with degenerate edge states crossing the Fermi level\cite{KaneMele} which are protected by time-reversal symmetry (TRS). Copies of the spin currents which propagate in opposite directions along the edges of the sample occur conserving TRS. Since the system also has inversion symmetry (IS) it is straightforward to characterize the topological properties of the system through the topological bulk invariant $Z_2$ using \eqref{nu}. The QSH has $\nu=1$ since the parity eigenvalues of the occupied bands at the TRIM are: $(\delta_\Gamma,\delta_{M_1},\delta_{M_2},\delta_{M_3})=(-1,-1,-1,1)$. Apart from the $Z_2$ invariant, we can obtain additional information about the topological properties of the QSH phase through the spin Chern numbers $c_{\uparrow n}$ ($-c_{\downarrow n}$) which can be unambiguously defined on isolated bands. Under the action of SOC, bands are strongly deformed eventually leading to the closing of the gap between two bands. At that point, the spin Chern numbers, $c_{n\uparrow}$($=-c_{\downarrow n}$), of the two bands involved in the closure, can change their values (see Fig. \ref{QSHBItransition}). Interestingly, we find that the spin Chern numbers obtained for $0<\lambda_{SO} <0.2t$: $c_\uparrow =(-1,1,-1,2,0,-1)$, change to $c_\uparrow =(-1,0,0,-1,3,-1)$ 
in the interval, $0.2 < \lambda_{SO} < 0.5$, 
due to the touching of the second and third bands at the $\Gamma$-point and of the fourth and fifth bands at the $K$-point before entering the SM phase. Despite these changes on the band spin Chern numbers, the total spin Chern number of the insulating QSH state at $f=1/2$ $C_\uparrow=c_{\uparrow 1}+c_{\uparrow 2}+c_{\uparrow 3}$ remains constant:  $C_\uparrow=-1$. 
In contrast, at $f=1/3$ the total spin Chern number changes from $C_\uparrow=0$ to $C_\uparrow=-1$ around $\lambda_{SO}\sim 0.2t$ signalling a transition from a band to a topological insulator.\cite{ruegg}
 
 \subsection{Topological semimetal}
 
 A SM phase was found in a previous analysis of model \eqref{Hamiltonian} at half-filling but with no Coulomb interaction, $U=0$.\cite{ruegg} Our phase diagram of Fig. \ref{HFSOCUdiagram} shows how a SM phase is stable in a broad range of $U, \lambda_{SO}$ values up to a substantial Coulomb interaction, $U>6t$. At weak coupling, a transition from the QSH phase to the SM phase occurs at a critical SOC of $\lambda_{SO}=0.5t$. While the system has a gap between the third (valence) and fourth (conduction) bands for $\lambda_{SO}< 0.5 t$, the Fermi level crosses both bands simultaneously for $\lambda_{SO}>0.5t$ as shown in Fig. \ref{QSHBItransition}. We identify the metallic phase as a SM due to the small overlap between the valence and conduction bands. Although this SM phase preserves TRS and IS we cannot associate a $Z_2$ invariant in the SM phase since it is gapless. However, this does not mean that the SM is topologically trivial. The SM phase hosts chiral currents in each spin sector similar but weaker than the chiral currents of the QSH as shown in Fig. \ref{Schemes}{\color{blue}(b)} which suggest non-trivial topology. We have investigated this possibility by computing spin Berry phases: $\gamma_{\uparrow n}$ ($-\gamma_{\uparrow n}$) associated with the Fermi surfaces crossing the $n$ partially filled bands. We find non-zero values of $\gamma_{\uparrow n}$ indicating a topological semimetallic state characterized by a non-zero spin Hall conductivity, $\sigma_{xy}^s$, which can be experimentally detected as discussed below in Sec. \ref{sec:exp}. Hence, we conclude that the present SM phase has non-trivial topological properties.

\begin{figure}[b!]
   \centering
   \includegraphics[width=2.8cm]{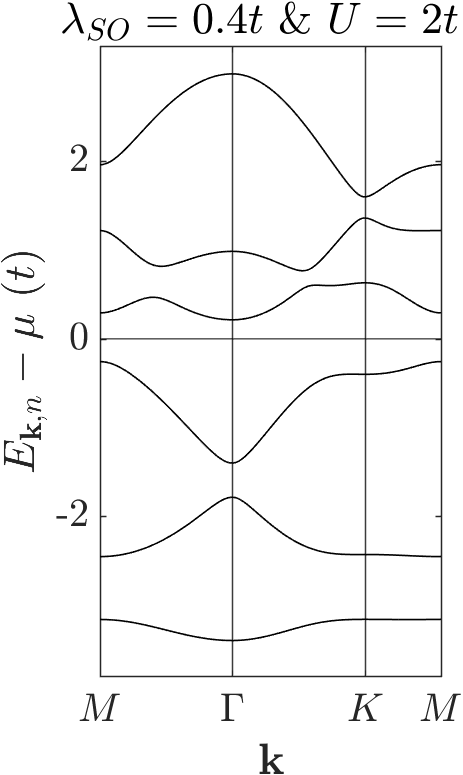}
   \hspace{0.8cm}
   \includegraphics[width=2.8cm]{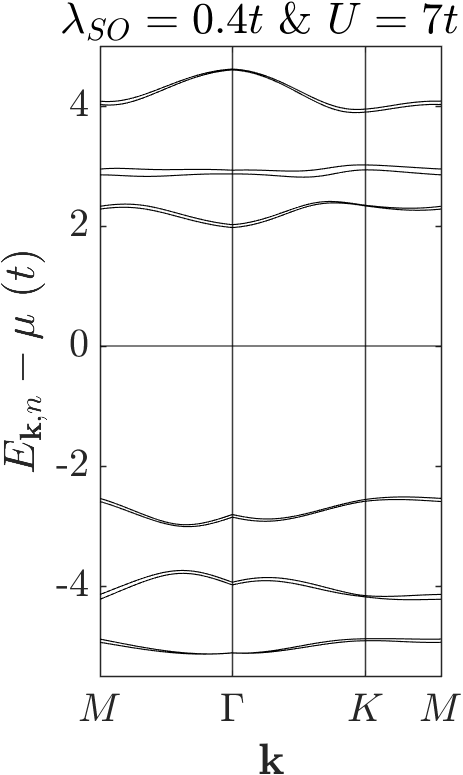}
   \includegraphics[width=2.8cm]{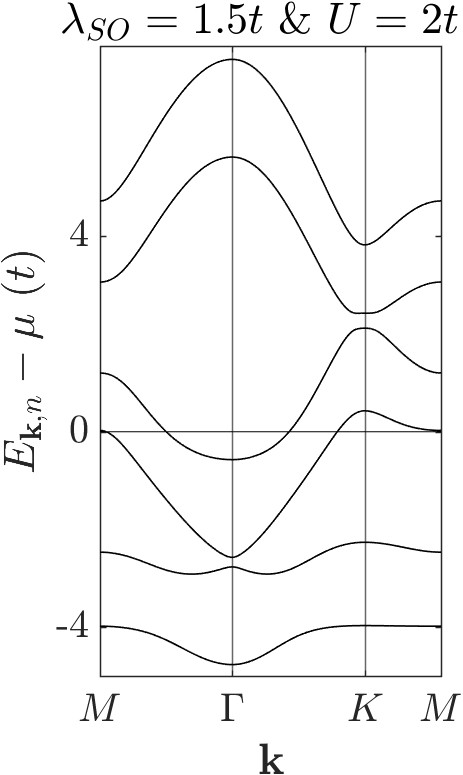}
   \hspace{0.8cm}
   \includegraphics[width=2.8cm]{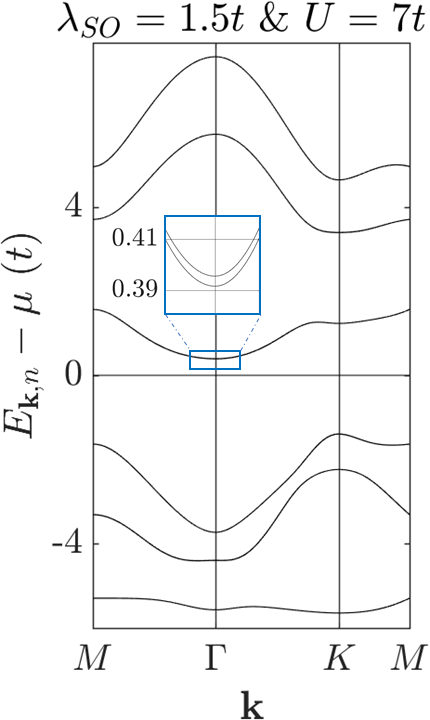}
       \caption{Hartree-Fock band dispersions across the QSH to AFI and the SM to $3/2$-MI phase. $U$ increases for $\lambda_{SO}=0.4t$ (top panels) and for $\lambda_{SO}=1.5t$ (bottom panels), respectively. (a) $\lambda_{SO}=0.4t$ and $U=2t$. In the QSH phase the bands are doubly degenerate in the whole FBZ due to the presence of TRS and IS. (b) $\lambda_{SO}=0.4t$ and $U=7t$. The bands are splitted by the spontaneous breaking of TRS due to magnetic order becoming an AFI. 
       (c) $\lambda_{SO}=1.5t$ and $U=2t$. We are in the SM region with both IS and TRS preserved. (d) $\lambda_{SO}=1.5t$ and $U=7t$. The gap is opened by the action of $U$ driving the system into the $3/2$-MI phase. Similarly the splitting takes place due to TRS breaking as can be observed in the inset.}
 \label{QSHAFMtransition}
 \end{figure}

\subsection{Antiferromagnetic insulating phase}
When the on-site Coulomb repulsion $U$ becomes sufficiently strong spin ordering occurs driving the system into a magnetic insulating phase. Our phase diagram of Fig. \ref{HFSOCUdiagram} shows how for $U > 2.5 t$ and SOC $0<\lambda_{SO}\approx 1.1t$ an AFI phase is stabilized. In this phase the spins are ordered according to the color pattern of Fig. \ref{Schemes}{\color{blue} (c)}, in which the inter-triangle n.n. spins are always opposite. This classical state is similar to the one found in a Heisenberg antiferromagnetic (HAFM) model with ${\bf q}=0$ on a $42$-sites cluster of the DHL in absence of SOC. The band structure for fixed $\lambda_{SO}=0.4t$ displayed in Fig. \ref{QSHAFMtransition}{\color{blue}(b)} shows how a band splitting proportional to $U$ occurs when the magnetic order 
order sets in disrupting the QSH phase. The breaking of TRS and IS does not allow the calculation of either the $Z_2$ invariant nor the spin Chern numbers $c_\uparrow$. This is because the spin sectors are mixed up by $U$ so that the spin Chern numbers become ill defined. In the AFI phase nonzero chiral bond currents $\langle c_{i \uparrow}^\dagger c_{j\uparrow} \rangle$, similar to the ones in the QSH phase arise but with
dominating n.n. bond compared to n.n.n. bond currents as Fig. \ref{Schemes}
shows. The presence of such chiral currents may indicate that the AFI phase does have nontrivial topological properties. 

\subsection{Magnetic insulator with effective $S=3/2$}
In the large $U$ and $\lambda_{SO}$ regime we find another type of magnetic insulator. All the spins of the electrons, $S_i^z$, in a  triangle are oriented in the same direction leading to an effective $S=3/2$ local moment per triangle. These moments arranged in an underlying honeycomb lattice order AF as Fig. \ref{Schemes}{\color{blue}(d)} shows. Thus, we call this magnetic state, effective $S=3/2$ magnetic insulator ($3/2$-MI). It can be rationalized from the fact that it is the only magnetic ordered state which allows AF alignment between all n.n.n. induced by the large SOC. This state breaks TRS and IS due to the emergence of local spontaneous magnetization with opposite direction in each triangle. Band splitting is observed although it is much smaller (see the inset of Fig. \ref{QSHAFMtransition}{\color{blue}(d)}) than in the AFI phase. This can be attributed to the much smaller Fock terms $\xi_i+\eta_i$ in the present
$S=3/2$-MI phase. However, close to the transition from the SM to $3/2$-MI there is an enhancement of the Fock terms making the band splitting huge in a small range of $U$ values. 

As found in the weakly interacting QSH and SM phases, the SOC induces spontaneous currents in the AFI and the $S=3/2$-MI phases as shown in Fig. \ref{Schemes} (c) and (d). While the n.n. currents in all the triangles of the AFI have the same direction, they have alternating directions in the two different triangles of the $S=3/2$-MI phase. We attribute this difference to the different local magnetization of the triangles in the two phases. On the other hand, the currents around the n.n.n. hexagonal loops are more prominent in the $3/2$-MI phase than in the AFI phase since the former phase needs a larger $\lambda_{SO}$ to be stabilized. The different amplitudes of the currents in the two triangles of the $3/2$-MI phase may also be associated to the interplay between SOC effects and the opposite local magnetization of the two triangles. Although the existence of the spontaneous currents do suggest topological features in the magnetic phases, a topological invariant different from the conventional ones (like the $Z_2$ or $C_\sigma$ used here for the non-interacting phases) should be introduced to characterize the 
possible non-trivial topology of these phases.

\section{Beyond mean-field theory: RVB quantum spin liquid vs. magnetic order}
\label{sec:ed}
It is interesting to go beyond mean-field theory analyzing possible electron correlation effects on the various phases of our HF phase diagram \ref{HFSOCUdiagram}, particularly on the large-$U$ magnetic phases found. Quantum fluctuations neglected in HF can distort and even destroy the classical-type magnetically ordered states found.  Here, we are particularly interested in the possibility that, at large-$U$, quantum spin liquid (QSL) phases may arise due to the frustration associated with the triangular coordination of the decorated honeycomb lattice. Applying ED techniques and RVB theory on our model \eqref{Hamiltonian} in small clusters we find that while an RVB state dominates at weak SOC, 
$\lambda_{SO}\ll t$ instead of the mean-field AFI, the magnetically ordered $3/2$-MI state found in HF survives to quantum fluctuations at large SOC $\lambda_{SO}\gg t$.

\subsection{Magnetic correlations on small clusters}
\label{sec:ed1}
\begin{figure}[t!]
		\centering
		\begin{center}
    	\includegraphics[width=4.25cm]{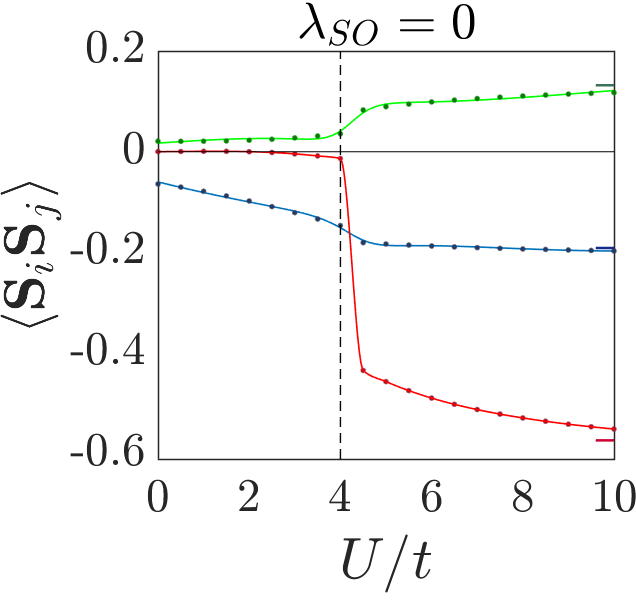}
    	\includegraphics[width=4.25cm]{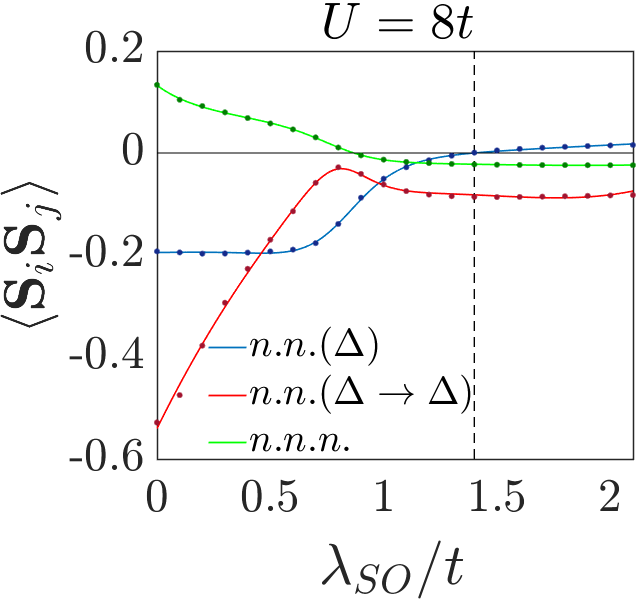}
		\end{center}
		\caption{Non-local spin correlations $\langle {\bf S_{i}S_{j}}\rangle$ in the Hubbard model with $U$ and $\lambda_{SO}$ on a six-sites cluster. The n.n. spin correlations $\langle {\bf S_{i}S_{j}}\rangle^{\bigtriangleup}$ and $\langle {\bf S_{i}S_{j}}\rangle^{\bigtriangleup\rightarrow\bigtriangleup}$ are represented in blue and red respectively, whereas the n.n.n $\langle {\bf S_{i}S_{j}}\rangle$ in green. On the left the spin correlations are displayed as a function of $U$ in absence of SOC ($\lambda_{SO}=0$). It can be observed a transition to a more correlated state at $U=4t$ marked as a dashed vertical line. The spin correlations got from RVB are also indicated as colored dashes. On the right the dependence of $\langle {\bf S_{i}S_{j}}\rangle$ with SOC at fixed $U=8t$ is shown. The dashed vertical line at $\lambda_{SO}=1.4t$ indicates when the $\langle {\bf S_{i}S_{j}}\rangle^{\bigtriangleup}$ becomes positive.
		}
 \label{SijU8}		
\end{figure}

We analyze the spin correlations obtained with ED on a small cluster with $N_s=6$ sites. In Fig. \ref{SijU8}{\color{blue}(a)} we show the dependence of spin correlations with $U$ and no SOC. At around $U=4t$ the magnitude of spin correlations displays a strong enhancement signalling short range spin ordering inside the cluster. While both kind of n.n. spin correlations are AF, the n.n.n. are FM. The spin correlations saturate rapidly with increasing $U$ to the $U\gg t$ values shown as colored dashes in  Fig. \ref{SijU8}{\color{blue}(a)} already at $U>(8-10)t$. 
It is worth pointing out how, above the transition point $U>4t$,
the magnitude of the inter-triangle n.n. spin correlations are much larger than the intra-triangle, {\it i. e.}, the ratio $r={\langle {\bf S_{i}S_{j}}\rangle^{\bigtriangleup\rightarrow\bigtriangleup} \over \langle {\bf S_{i}S_{j}}\rangle^{\bigtriangleup}} >1$ despite the fact that all n.n. hoppings (inter and intra-triangle) are the same. The anisotropy in the spin correlations becomes substantial, $r \rightarrow 3.5$, in the $U/t \rightarrow \infty$ limit. We show below how an RVB ground state of our six site cluster provides a faithful description of the exact ground state of the cluster recovering naturally such large unexpected anisotropy in the n.n. spin correlations. 

We also explore the AFI to $3/2$-MI transition induced by SOC
found in the HF analysis by computing the dependence of the spin correlations on $\lambda_{SO}$ at a large $U=8t$ as Fig. \ref{SijU8}{\color{blue}(b)} shows. For small $0<\lambda_{SO}\sim 1.4 t$ both n.n. intra-triangle and n.n. inter-triangle spin correlations, $\langle {\bf S_{i}S_{j}}\rangle^{\bigtriangleup\rightarrow\bigtriangleup}$, remain AF as for the case with no SOC. At a larger SOC, $\lambda_{SO} > 1.4t$, $\langle {\bf S_{i}S_{j}}\rangle^{\bigtriangleup}$ becomes positive indicating a FM coupling between spins within the same triangle, while $\langle {\bf S_{i}S_{j}}\rangle^{\bigtriangleup\rightarrow\bigtriangleup}$ remains AF. These spin correlations are consistent with the $3/2$-MI state obtained from HF theory shown in Fig. \ref{Schemes}{\color{blue}(d)}. 

\begin{figure}[t!]
		\centering
		\begin{center}
    	\includegraphics[width=4.25cm]{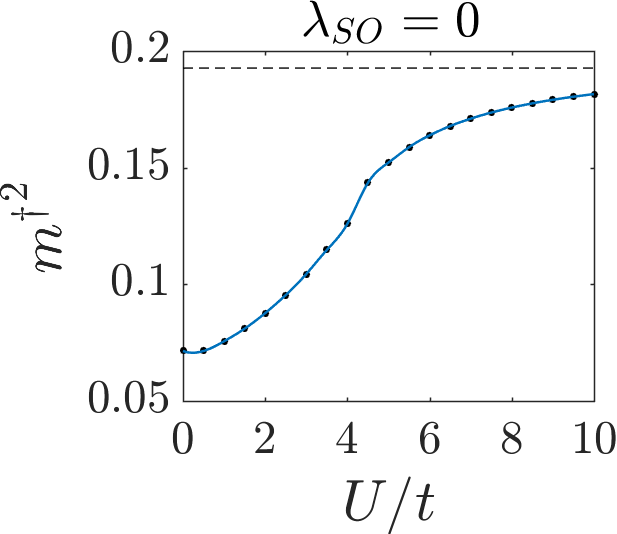}
		\includegraphics[width=4.25cm]{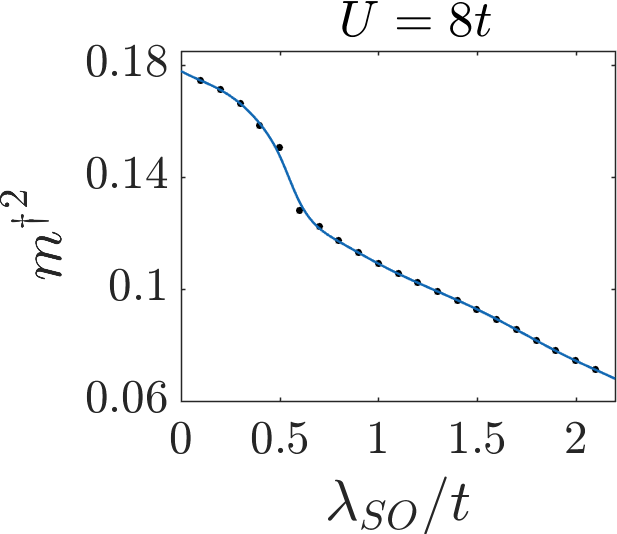}
		\end{center}
		\caption{Dependence of the magnetic order parameter ${m^+}^2$ 
		on $U$ and SOC on $N_s=12$ clusters. On the left panel ${m^+}^2$ is represented as a function of $U$ for fixed $\lambda_{SO}=0$. The dashed horizontal line shows the saturation for large $U$.  
		In the same way, the right panel shows the dependence of the order parameter on $\lambda_{SO}$ at fixed $U=8t$.}
 \label{orderparameters}		
\end{figure}

Further insight into the various transitions found can be obtained
by analyzing the $U$ dependence of the magnetic order parameter, $m^\dagger$, introduced earlier\cite{Heisenberg}: 
\begin{equation}
    m^{+^2}=\frac{1}{N_s^2}\sum_{i,j}\abs{\langle {\bf S_{i}S_{j}}\rangle},
\end{equation}
where $N_s$ is the number of sites of the cluster. In Fig. \ref{orderparameters}{\color{blue}(a)} we show the dependence of $m^{+^2}$ in the Hubbard model on a $12$-site cluster in absence of SOC: $\lambda_{SO}=0$. The rapid increase of $m^{+^2}$ with $U$ indicates the building up of AF spin correlations which start to saturate around $U \approx (6-7)t$ signalling the formation of a 
$S=1/2$ state with short range magnetism. In order to make contact with the Heisenberg model we show in Fig. \ref{orderparameters}{\color{blue}(a)} the value of $m^{+^2}=0.1927$ obtained for $U\gg t$ and no SOC, which is consistent with the order parameter $m^{\dagger^2}\sim 0.2167$ extrapolated to a 12-site cluster.\cite{Heisenberg} These values are strongly suppressed in the extrapolation to the thermodynamic limit of the Heisenberg model which find $m^{+^2} \approx 0.0025$ which, at the same time, is much smaller than the classical value, $m^{+^2} \approx 0.1665$, indicating a spin disordered state. The SOC dependence of $m^{+^2}$ at fixed $U=8t$ is shown in Fig. \ref{orderparameters}{\color{blue}(b)}. The $m^{+^2}$ is suppressed with SOC as expected from the suppression of the spin correlations $\langle {\bf S_{i}S_{j}}\rangle$ shown in Fig. \ref{SijU8}{\color{blue}(b)} which is associated with the formation of the $3/2$-MI phase. 

Hence, our analysis suggests that a transition from a quantum paramagnet to a quantum spin state with short range magnetic order occurs for $U > 5t$ at weak SOC. Increasing SOC drives this state into a $3/2$-MI state around a critical $\lambda_{SO}=1.4t$. This picture is qualitatively consistent with the HF phase diagram with, however, the AFI state replaced by a spin disordered phase. 

\subsection{RVB state at large $U$ and weak SOC}
\label{sec:ed2}
Previous studies of the $S=1/2$ Heisenberg model on the decorated honeycomb lattice indicate the existence of a valence bond crystal (VBC) state with no long range magnetic order.\cite{Heisenberg,Orus} The destruction of magnetic order can be associated with the triangular coordination of the 
lattice inducing strong magnetic frustration. An interesting 
property of such VBC is the fact that n.n. inter-triangle spin correlations are much stronger than intra-triangle despite the fact that all n.n. exchange couplings are the same. This was already found in previous ED studies of the $S=1/2$ Heisenberg model on the decorated honeycomb lattice on clusters up to $N_s=42$ sites.\cite{Heisenberg} In order to compare with these results we fix $U=100t$ and $\lambda_{SO}=0$ in our Hubbard model on a $N_s=6$ cluster, finding that indeed the AF spin correlations are very different: $\langle {\bf S}_i {\bf S}_j \rangle ^{\bigtriangleup\rightarrow\bigtriangleup}=-0.600$
and $\langle {\bf S}_i {\bf S}_j \rangle^{\bigtriangleup}=-0.208$.
Our results are consistent with the spin correlations in the Heisenberg model obtained through ED on much larger clusters of $N_s=42$ sites: $\langle {\bf S}_i {\bf S}_j \rangle ^{\bigtriangleup\rightarrow\bigtriangleup}=-0.591$ and $\langle {\bf S}_i {\bf S}_j^{\bigtriangleup} \rangle=-0.168 $.\cite{Heisenberg} The good agreement between the n.n. spin correlations of $N_s=6$ and $N_s=42$ clusters indicates that the short range AF correlations of isolated triangular dimers are dominant. This is further corroborated by the large singlet-triplet gap, $\Delta\approx 0.38J$, and the lack of spin singlet excitations within the gap weakly dependent on the cluster size\cite{Heisenberg} in close  resemblance with the spin excitation spectrum of isolated triangular dimer units. This motivates an analysis of triangular dimers in order to gain insight about the ground state of the system in the thermodynamic limit when $U\gg t$. 

\begin{figure}[t!]
    \centering
    \subfigure[]{\includegraphics[width=1.7cm]{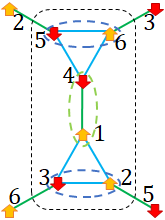}}    
    \hspace{0.3cm}    
    \subfigure[]{\includegraphics[width=1.7cm]{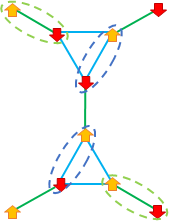}}    
    \hspace{0.3cm}    
    \subfigure[]{\includegraphics[width=1.7cm]{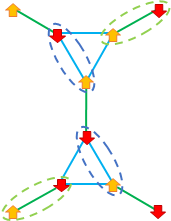}}  
    \hspace{0.3cm}
    \subfigure[]{\includegraphics[width=1.7cm]{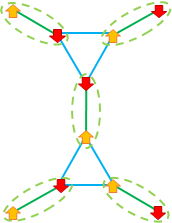}}      
    \caption{Valence bond configurations considered in the RVB wavefunction
    on a six-sites cluster. The singlet pairings between n.n. spins composing each valence bond are marked with ellipses: intra-triangle singlets (blue dashed) and inter-triangle (green dashed). The RVB state consisting on a linear combination of these four configurations provides a faithful description of the ground state of the $S=1/2$ Heisenberg model on a six-site cluster of the decorated honeycomb lattice.
    }
    \label{QuantumSchemes}		
\end{figure}

We consider an RVB state as a possible candidate for the ground state wavefunction of the Heisenberg model on a DHL. Such RVB is constructed as a linear combination of all possible configurations in which n.n. spins are paired up into singlets. On a six-site cluster, the RVB state consists on the linear combination of the four possible valence bond (VB) configurations of Fig. \ref{QuantumSchemes}, which can be expressed as:
\begin{align}
\ket{RVB}&=(14)(23)(56)+(13)(25)(46)\nonumber\\
&+(12)(36)(45)-(14)(25)(36),
\label{eq:rvb}
\end{align}
where $(ij)= {1 \over \sqrt{2}} \left( c^\dagger_{i\uparrow} c^\dagger_{j\downarrow} - c^\dagger_{i\downarrow} c^\dagger_{j\uparrow} \right) |0\rangle $ is a singlet between n.n. sites $ij$ with the numeration 
of the sites as shown in Fig. \ref{QuantumSchemes} (see the Appendix \ref{sec:RVB} for further details). The overlap between this RVB and 
the exact ground state wavefunction obtained from ED for $U\gg t$ is nearly one: $\langle RVB |\Psi_0 \rangle=0.9988$, and the RVB energy
is only a $0.16\%$ higher than the exact ground state energy obtained
with ED. These facts indicate that such RVB state provides a very good
description of the exact ground state of the Heisenberg model on this six-site cluster including the large difference between the n.n. inter- and intra-triangle spin correlations is naturally captured by such RVB state. Indeed, we find that while $\langle {\bf S_{i}S_{j}}\rangle^{\bigtriangleup}=-{13 \over 17 \times 4} \approx -0.191$, $\langle {\bf S_{i}S_{j}}\rangle^{\bigtriangleup\rightarrow\bigtriangleup}=-{39 \over 17 \times 4} \approx -0.573$ in very good agreement with the exact result. This allows to understand the large anisotropy in the n.n. spin correlations arising in the Heisenberg model on the DHL despite isotropic n.n. exchange couplings. This canbe understtod as a consequence of the interference of the VB configurations contributing to the RVB wavefunction. Note also that the RVB configurations involving singlets between triangles occur with an opposite sign to the configurations involving singlets in the triangles.
From the above analysis, we conclude that the RVB state is a good candidate for the ground state wavefunction of the Heisenberg model on the infinite DHL. Such RVB state would read:
\begin{align}
\ket{RVB}&=\sum_{m} a_m \ket{m},
\label{eq:rvb1}
\end{align}
where $\ket{m}$ denotes a VB configuration in which all 
n.n. spins of the DHL are paired up into singlets and 
$a_m=\pm 1$. Although this RVB state neglects 
VB configurations containing singlets between spins at sites beyond the n.n., we expect that it will still provide a good description of the n.n. spin correlations of the exact ground state including the unexpected large anisotropy in the n.n. spin 
correlations.
\subsection{Magnetic order at large $U$ and strong SOC}
\label{sec:ed3}
\begin{figure}[t!]
		\centering
		\centering
        \includegraphics[width=6cm]{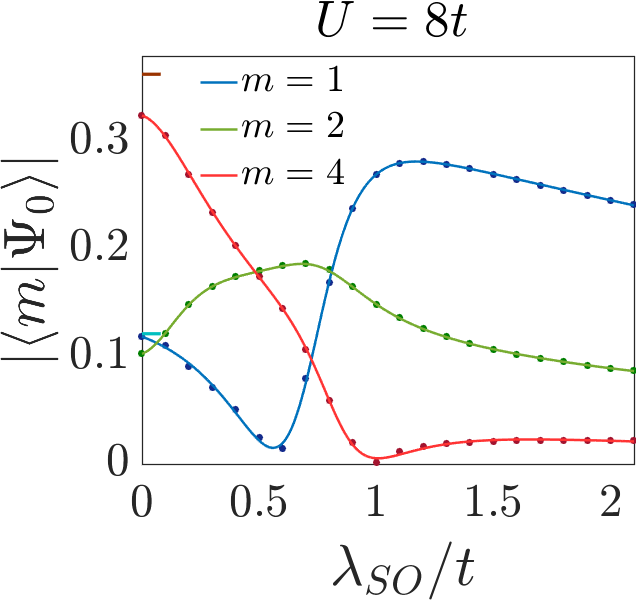}
		\caption{Dependence of ground state wavefunction on SOC. The modulus of the coefficients of the dominant non-ionic configurations $\abs{\bra{m}\Psi_{0}\rangle}$ for $U=8t$ labeled as in the RVB wavefunction (see Appendix \ref{sec:RVB}) are shown. The SOC, $\lambda_{SO}$, favours the formation of a $S=3/2$ AF configuration in which each triangle contains three parallel spins which are oppositely oriented to the spins of its n.n. triangles as shown in Fig. \ref{Schemes}{\color{blue}(d)}. The values of these three coefficients of the RVB model ($U\gg t$) at $\lambda_{SO}=0$ are displayed as colored dashes: $\bra{1}RVB\rangle =\bra{2}RVB\rangle=0.121$ in cian and $\bra{4}RVB\rangle =0.364$ in dark red \eqref{eq:developedrvb}.
		}
 \label{coefficients}		
\end{figure}

We can gain further insight into the formation of the $3/2$-MI state at strong SOC from a direct inspection of the exact ground state wavefunction, $\ket{\Psi_0}$, on the $N_s=6$ site cluster. This provides useful complementary information to the spin correlations discussed previously. In Fig. \ref{coefficients}, the
dependence of the exact wavefunction coefficients: $|\bra{m}\ket{\Psi_0}|$ on $\lambda_{SO}$ is shown at a large $U=8t$. The three dominant configurations, $\ket{m}$, plotted are the non-ionic configurations of the wavefunction and are given explicitly in Appendix \ref{sec:RVB}. While for $\lambda_{SO} \rightarrow 0$ the wavefunction coefficients match almost perfectly with the RVB state ones shown as colored dashes, in the $\lambda_{SO}\gg t$ limit, the $\ket{1}=c^\dagger_{1\uparrow}c^\dagger_{2\uparrow}c^\dagger_{3\uparrow}c^\dagger_{4\downarrow}c^\dagger_{5\downarrow}c^\dagger_{6\downarrow}|0\rangle$ and $\ket{20}=c^\dagger_{1\downarrow}c^\dagger_{2\downarrow}c^\dagger_{3\downarrow}c^\dagger_{4\uparrow}c^\dagger_{5\uparrow}c^\dagger_{6\uparrow}|0\rangle$ configurations dominate the wavefunction. These two configurations, 
are equivalent energetically having the same weight in the wavefunction of our finite cluster. However, in the thermodynamic limit we expect that only one of these configurations is picked up due to the spontaneous symmetry breaking process leading to the classical $3/2$-MI phase of Fig. \ref{Schemes}{\color{blue}(d)}. The magnetic moment of such magnetic state would be effectively decreased, {\it i.e.} $S\lesssim 3/2$, by quantum fluctuations coming from the non-negligible weight of configuration $\ket{2}$ in the ground state. More careful work on larger clusters is needed to confirm our prediction.

By comparing both quantities $\langle {\bf S}_i \cdot {\bf S}_j\rangle$ and $\abs{\bra{m}\Psi_{0}\rangle}$ in Fig. \ref{SijU8}\textcolor{blue}{(b)} and Fig. \ref{coefficients} respectively, we note that the critical SOC at which the $n.n.(\bigtriangleup)$ spin correlations become positive occurs at $\lambda_{SO}=1.4t$ which is somewhat larger than the value of $\lambda_{SO}=0.8t$ at which the $\ket{1},\ket{20} $ configurations start to dominate. This mismatch can be attributed to the small but non-negligible contributions of the ionic configurations still present at $U=8t$. 

We finally provide a simple explanation for the formation 
of the $3/2$-MI state. SOC acts as a chiral imaginary hopping, $\pm i\lambda_{SO}$, connecting n.n.n. sites. In the large-$U$ limit: $U \gg \lambda_{SO}$ and such hopping leads to an AF spin exchange coupling, $J'= 4{\lambda_{SO}^2 \over U}$, between n.n.n. sites. However, due to the particular form of the SOC hopping term, it leads, instead to a coupling which differs somewhat from the pure Heisenberg type\cite{lehur}:
\begin{equation}
\mathcal{H}_{J'}=J'\sum_{\langle \langle ij \rangle \rangle} \left( -S_i^x S_j^x -S_i^y S_j^y +S_i^z \right)
\end{equation}
This model couples ferromagnetically the $x, y$-components and antiferromagnetically the $z$ components of the n.n.n spins of the lattice. Thus, the complete effective spin model arising from the original Hubbard model in the $U\gg\lambda_{SO}, t$ limit reads:
\begin{equation}
\mathcal{H}_{J-J'}=J\sum_{\langle ij \rangle} {\bf S}_i {\bf S}_j+H_{J'},
\label{hjjp}
\end{equation}
where $J= 4 t^2/U$. Hence, we can expect that in the limit, $\lambda_{SO}\gg t$,  $J'\gg J$ so that it is energetically favorable to orientate the $z$-component of two n.n.n. spins of the DHL in opposite directions while keeping the components in the $x,y$ plane aligned. This is achieved if the $z$-component of the three spins inside a triangle are FM aligned and AFM aligned with the three spins in the n.n. triangle. A transition from the RVB to a $3/2$-MI state is found around  a critical SOC of $\lambda_{SO}\sim 0.8t$, as discussed in appendix 
\ref{sec:RVB2} based on this $J-J'$ model. This critical $\lambda_{SO}$ is 
in good agreement with our results on the Hubbard model for $U=8t$ shown in Fig. 5(b) indicating that the large-$U$ regime has been reached. 
\section{Spin Hall effect}
\label{sec:exp}
\begin{figure}[t!]
		\centering
		\begin{center}
			\includegraphics[width=6.18cm]{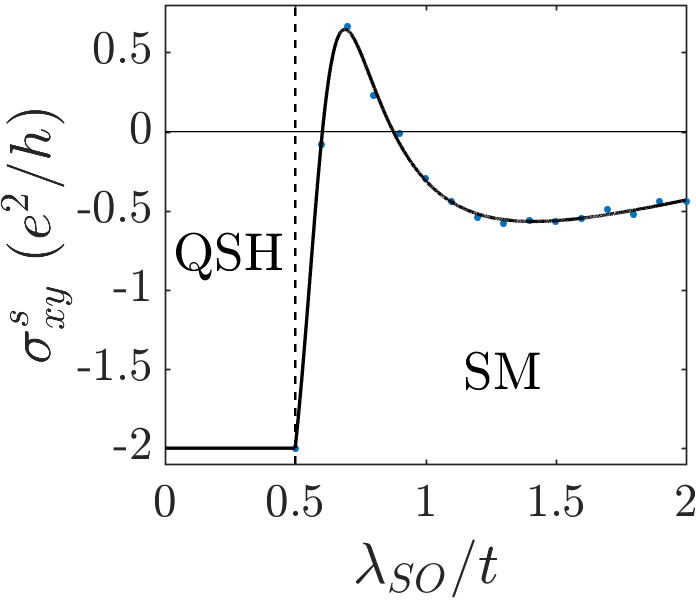}
		\end{center}
		\label{gamma1}
		\caption{Spin Hall conductivity $\sigma_{xy}^s$ as a function of $\lambda_{SO}$ at $U=0$. The dashed vertical line indicates the point at which the system turns into SM. We find nonzero conductivity inside the SM zone keeping the characteristic chiral currents of the QSH phase as can be seen in Fig. \ref{Schemes}{\color{blue}{(b)}}.
		} 
 \label{spinHallconductivity}		
\end{figure}

We now discuss the implications of some of our results on experiments.
We consider the spin Hall conductivity, $\sigma^s_{xy}$,  in the non-interacting limit of the model. We are particularly interested in the dependence of $\sigma^s_{xy}$ with increasing $\lambda_{SO}$. 
In the QSH phase obtained at $0<\lambda_{SO}<0.5 t$ the spin Hall conductivity can be obtained from the spin Chern numbers of the occupied bands as: $\sigma_{xy}^s=-\frac{e^2}{h} \sum_{n=1}^{N_c}(c_{\uparrow n}-c_{\downarrow n})$ where $N_c$ denotes the highest occupied band
of the system. For $\lambda_{SO}>0.5t$, in the SM phase we can evaluate the spin Hall conductivity based on the Haldane expression for 2D metallic systems\cite{Haldane}: 
\begin{equation}
\sigma_{xy}^s=-\frac{e^2}{h} \sum_{n=1}^{N_c}(c_{\uparrow n}-c_{\downarrow n})- \frac{e^2}{h}\sum_{n=N_{c}+1}^{N_c+2} {(\gamma_{\uparrow n}-\gamma_{\downarrow n}) \over 2 \pi}
\label{condformula}
\end{equation}
which involves the spin Chern numbers, $c_{\sigma,n}$, of the occupied bands and the spin Berry phases, $\gamma_{\sigma n}$ of the closed Fermi surface sections associated with the bands crossing the Fermi energy. The $n=N_c+1$ and $n=N_c+2$ bands are the two partially filled bands in the SM phase. The calculated spin Hall conductivity as a function of $\lambda_{SO}$ is shown in Fig. \ref{spinHallconductivity}. For $0<\lambda_{SO}<0.5$
the system is in the QSH and $\sigma_{xy}^s$ is quantized with a value of $\sigma_{xy}^s=-2 e^2/h$. When $\lambda_{SO}>0.5t$ the system enters the SM phase and a strong variation of $\sigma_{xy}^s$ with $\lambda_{SO}$ occurs. It is interesting to notice that the spin Hall conductivity, although non-quantized is non-zero in the SM phase.  

The behaviour of $\sigma_{xy}^s$ with $\lambda_{SO}$ observed in Fig. \ref{spinHallconductivity} is associated with changes in the spin Chern numbers due to band gap closings induced by SOC as can be observed in Fig. \ref{QSHBItransition}. These changes, in turn, influence the spin Berry phases ($\gamma_{\sigma n}$) on the Fermi surface which are fractions of the spin Chern numbers associated with the partially filled bands of the SM phase.
Based on this observation we can explain the maximum in $\sigma_{xy}^s$ around $\lambda_{SO} \approx 0.7t$. For this
SOC, the third and fourth bands touch at the $M$-points so that the spin Chern numbers become: $c_\uparrow =(-1,0,3,-4,1,1)$,
and the third (fourth) band reach their largest values found, $c_{\uparrow 3}=3$ ($c_{\uparrow 4}=-4$).
Spin Hall conductivity experiments on DHL materials 
maybe the most direct way to probe the presence of the SM phase
found here. 
\section{Conclusions}
\label{sec:concl} 
The interplay of Coulomb repulsion and SOC on decorated honeycomb lattices (DHL) leads to a rich phase diagram including topological insulating and metallic phases as well as quantum spin liquid and magnetically ordered phases. 
 
At weak Coulomb repulsion and non-zero SOC, we find that the topological QSH and SM phases are stable up to moderate values of $U$. Our HF analysis predicts that the spontaneous counter-propagating chiral currents expected in the QSH persist in the SM phase as shown in Fig. \ref{Schemes}{\color{blue}(b)}. The non-zero spin Chern numbers obtained in the QSH leads to a quantized spin Hall conductivity (as expected in $Z_2$ topological insulators) of $\sigma^s_{xy}=-2{e^2 \over h}$. In contrast, the SM phase is characterized by a finite but non-quantized $\sigma^s_{xy}$ arising from the non-zero spin Chern numbers of the occupied bands as well as from the spin Berry phases of the Fermi surfaces. Hence, measuring a non-zero spin Hall conductivity would provide evidence for such topological metallic phase.

Magnetic ordering occurs at a critical $U_c(\lambda_{SO})$ obtained from HF that increases with $\lambda_{SO}$. This increase is attributed to the deformation of the flat band at the Fermi energy\cite{manuel} with no SOC making the QSH and SM phases more unstable to magnetic order as $\lambda_{SO}$ is decreased. Two different spin ordered states emerge depending on the values of SOC: an AFI phase for  $0<\lambda_{SO}<1.1$ and an effective $S=3/2$ magnetic insulator, the $3/2$-MI phase at larger SOC.

In order to explore the robustness of the HF phases for strong Coulomb repulsion we have analyzed electron correlations effects based on ED techniques. At weak SOC we conclude that the ground state of the Hubbard model on the DHL is an RVB state in contrast to the classical AFI phase found with HF. RVB theory on small clusters naturally explains the unexpected large anisotropy of the n.n. spin correlations noted previously.\cite{Heisenberg,Orus} A VBS is proposed as the ground state of the Heisenberg the model on the DHL since the n.n. inter-triangle, $(\bigtriangleup\rightarrow\bigtriangleup)$, spin correlations are 3.5 times larger than the intra-triangle, $(\bigtriangleup)$, in spite of the isotropic n.n. exchange couplings considered. Here, we find that such anisotropy arises naturally from the interference effects encoded in the RVB state. On the other hand, our ED analysis on small clusters indicates that a FM alignment of the spins in each triangle occurs consistent with the classical $3/2$-MI state found in HF. This can be easily understood in terms of a $J-J'$ Heisenberg model in which the AF n.n.n. exchange coupling, $J' \propto \lambda_{SO}^2/U>0$, induced by SOC wins over the $n.n.$ exchange, $J \propto t^2/U>0$, at large SOC, {\it i. e.}, $J'\gg J$. In this situation, it is energetically favourable to align FM the three spins in a given triangle and AF with the three spins in any n.n. triangle leading to a non-
fully saturated $3/2$-MI state.
  
Our results open interesting avenues including the possibility of inducing superconductivity in the DHL by doping the RVB state found at half-filling
as proposed for the cuprates.\cite{anderson87} On the other hand, the DHL at strong Coulomb repulsion and strong SOC provides a playground to study quantum magnetism of AF coupled $S=3/2$ local moments in a honeycomb lattice. These issues are left for future studies.

\acknowledgments
We acknowledge financial support from Spanish Ministry of Science and Innovation (RTI2018-098452-B-I00) MINECO/FEDER, Uni\'on Europea and through the “Mar\'ia de Maeztu” Programme for Units of Excellence 
in R\&D (CEX2018-000805-M).
\appendix

\section{Heisenberg model and the RVB state on small clusters}
\label{sec:RVB}
Here we describe the similarities between the exact ground state obtained by ED on six-site clusters and the approximate RVB theory on small clusters. In the limit of $U \gg t$, our Hubbard model with SOC can be mapped onto a Heisenberg-type model on the decorated honeycomb lattice:
\begin{equation}
\mathcal{H}_{J-J'}=J\sum_{\langle ij  \rangle } {\bf S}_i {\bf S}_j+J'\sum_{\langle \langle ij \rangle \rangle} \left( -S_i^x S_j^x -S_i^y S_j^y +S_i^z S_j^z \right),
\label{hjjp1}
\end{equation}
neglecting constant terms. The first term describes a n.n. AF Heisenberg exchange coupling with $J=4t^2/U$
while the second term describes the magnetic exchange 
between n.n.n. induced by SOC, $J'=4\lambda_{SO}^2/U$.
This model contains frustration of spin order in 
the $z$ components due to competition between the n.n. 
AF $J$ and the n.n.n. AF $J'$. 
Also the x-y coupling between spins displays competition
between the n.n. AF $J$ and the n.n.n. FM $J'$. 

On the six-site cluster of Fig. \ref{QuantumSchemes}, this model can be expressed on the valence bond (VB) basis leading to a reduced $20 \times 20$ hamiltonian matrix instead of the $ 400 \times 400 $ of the Hubbard model. These VB basis states are:
\begin{eqnarray}
|1\rangle&=&c^\dagger_{1\uparrow}c^\dagger_{2\uparrow}c^\dagger_{3\uparrow}c^\dagger_{4\downarrow}c^\dagger_{5\downarrow}c^\dagger_{6\downarrow}|0\rangle
\nonumber \\
|2\rangle&=&c^\dagger_{1\uparrow}c^\dagger_{2\uparrow}c^\dagger_{3\downarrow}c^\dagger_{4\uparrow}c^\dagger_{5\downarrow}c^\dagger_{6\downarrow}|0\rangle
\nonumber \\
|3\rangle&=&c^\dagger_{1\uparrow}c^\dagger_{2\uparrow}c^\dagger_{3\downarrow}c^\dagger_{4\downarrow}c^\dagger_{5\uparrow}c^\dagger_{6\downarrow}|0\rangle
\nonumber \\
|4\rangle&=&c^\dagger_{1\uparrow}c^\dagger_{2\uparrow}c^\dagger_{3\downarrow}c^\dagger_{4\downarrow}c^\dagger_{5\downarrow}c^\dagger_{6\uparrow}|0\rangle
\nonumber \\
|5\rangle&=&c^\dagger_{1\uparrow}c^\dagger_{2\downarrow}c^\dagger_{3\uparrow}c^\dagger_{4\uparrow}c^\dagger_{5\downarrow}c^\dagger_{6\downarrow}|0\rangle
\nonumber \\
|6\rangle&=&c^\dagger_{1\uparrow}c^\dagger_{2\downarrow}c^\dagger_{3\uparrow}c^\dagger_{4\downarrow}c^\dagger_{5\uparrow}c^\dagger_{6\downarrow}|0\rangle
\nonumber \\
|7\rangle&=&c^\dagger_{1\uparrow}c^\dagger_{2\downarrow}c^\dagger_{3\uparrow}c^\dagger_{4\downarrow}c^\dagger_{5\downarrow}c^\dagger_{6\uparrow}|0\rangle
\nonumber \\
|8\rangle&=&c^\dagger_{1\uparrow}c^\dagger_{2\downarrow}c^\dagger_{3\downarrow}c^\dagger_{4\uparrow}c^\dagger_{5\uparrow}c^\dagger_{6\downarrow}|0\rangle
\nonumber \\
|9\rangle&=&c^\dagger_{1\uparrow}c^\dagger_{2\downarrow}c^\dagger_{3\downarrow}c^\dagger_{4\uparrow}c^\dagger_{5\downarrow}c^\dagger_{6\uparrow}|0\rangle
\nonumber \\
|10\rangle&=&c^\dagger_{1\uparrow}c^\dagger_{2\downarrow}c^\dagger_{3\downarrow}c^\dagger_{4\downarrow}c^\dagger_{5\uparrow}c^\dagger_{6\uparrow}|0\rangle
\nonumber \\
|11\rangle&=&c^\dagger_{1\downarrow}c^\dagger_{2\uparrow}c^\dagger_{3\uparrow}c^\dagger_{4\uparrow}c^\dagger_{5\downarrow}c^\dagger_{6\downarrow}|0\rangle
\nonumber \\
|12\rangle&=&c^\dagger_{1\downarrow}c^\dagger_{2\uparrow}c^\dagger_{3\uparrow}c^\dagger_{4\downarrow}c^\dagger_{5\uparrow}c^\dagger_{6\downarrow}|0\rangle
\nonumber \\
|13\rangle&=&c^\dagger_{1\downarrow}c^\dagger_{2\uparrow}c^\dagger_{3\uparrow}c^\dagger_{4\downarrow}c^\dagger_{5\downarrow}c^\dagger_{6\uparrow}|0\rangle
\nonumber \\
|14\rangle&=&c^\dagger_{1\downarrow}c^\dagger_{2\uparrow}c^\dagger_{3\downarrow}c^\dagger_{4\uparrow}c^\dagger_{5\uparrow}c^\dagger_{6\downarrow}|0\rangle
\nonumber \\
|15\rangle&=&c^\dagger_{1\downarrow}c^\dagger_{2\uparrow}c^\dagger_{3\downarrow}c^\dagger_{4\uparrow}c^\dagger_{5\downarrow}c^\dagger_{6\uparrow}|0\rangle
\nonumber \\
|16\rangle&=&c^\dagger_{1\downarrow}c^\dagger_{2\uparrow}c^\dagger_{3\downarrow}c^\dagger_{4\downarrow}c^\dagger_{5\uparrow}c^\dagger_{6\uparrow}|0\rangle
\nonumber \\
|17\rangle&=&c^\dagger_{1\downarrow}c^\dagger_{2\downarrow}c^\dagger_{3\uparrow}c^\dagger_{4\uparrow}c^\dagger_{5\uparrow}c^\dagger_{6\downarrow}|0\rangle
\nonumber \\
|18\rangle&=&c^\dagger_{1\downarrow}c^\dagger_{2\downarrow}c^\dagger_{3\uparrow}c^\dagger_{4\uparrow}c^\dagger_{5\downarrow}c^\dagger_{6\uparrow}|0\rangle
\nonumber \\
|19\rangle&=&c^\dagger_{1\downarrow}c^\dagger_{2\downarrow}c^\dagger_{3\uparrow}c^\dagger_{4\downarrow}c^\dagger_{5\uparrow}c^\dagger_{6\uparrow}|0\rangle
\nonumber \\
|20\rangle&=&c^\dagger_{1\downarrow}c^\dagger_{2\downarrow}c^\dagger_{3\downarrow}c^\dagger_{4\uparrow}c^\dagger_{5\uparrow}c^\dagger_{6\uparrow}|0\rangle.
\label{eq:config}
\end{eqnarray}
where $\ket{0}$ is the vacuum state of the 6-sites DHL. The hamiltonian $\mathcal{H}_{J-J'}=\mathcal{H}_{J}+\mathcal{H}_{J'}$ in this basis reads:
\begin{widetext}
\begin{align}
\mathcal{H}_{J}=\frac{J}{4}\left(\begin{array}{cccccccccccccccccccc} 3 & 0 & 0 & 2 & 0 & 2 & 0 & 0 & 0 & 0 & 2 & 0 & 0 & 0 & 0 & 0 & 0 & 0 & 0 & 0\\ 0 & -1 & 2 & 2 & 2 & 0 & 0 & 2 & 0 & 0 & 2 & 0 & 0 & 0 & 0 & 0 & 0 & 0 & 0 & 0\\ 0 & 2 & -1 & 2 & 0 & 2 & 0 & 0 & 0 & 0 & 0 & 2 & 0 & 2 & 0 & 0 & 0 & 0 & 0 & 0\\ 2 & 2 & 2 & -5 & 0 & 0 & 2 & 0 & 0 & 2 & 0 & 0 & 2 & 0 & 2 & 0 & 0 & 0 & 0 & 0\\ 0 & 2 & 0 & 0 & -1 & 2 & 2 & 0 & 2 & 0 & 2 & 0 & 0 & 0 & 0 & 0 & 0 & 0 & 0 & 0\\ 2 & 0 & 2 & 0 & 2 & -5 & 2 & 0 & 0 & 2 & 0 & 2 & 0 & 0 & 0 & 0 & 2 & 0 & 0 & 0\\ 0 & 0 & 0 & 2 & 2 & 2 & -1 & 0 & 0 & 0 & 0 & 0 & 2 & 0 & 0 & 0 & 0 & 2 & 0 & 0\\ 0 & 2 & 0 & 0 & 0 & 0 & 0 & -1 & 2 & 2 & 0 & 0 & 0 & 2 & 0 & 0 & 2 & 0 & 0 & 0\\ 0 & 0 & 0 & 0 & 2 & 0 & 0 & 2 & -1 & 2 & 0 & 0 & 0 & 0 & 2 & 0 & 0 & 2 & 0 & 0\\ 0 & 0 & 0 & 2 & 0 & 2 & 0 & 2 & 2 & -5 & 0 & 0 & 0 & 0 & 0 & 2 & 0 & 0 & 2 & 2\\ 2 & 2 & 0 & 0 & 2 & 0 & 0 & 0 & 0 & 0 & -5 & 2 & 2 & 0 & 2 & 0 & 2 & 0 & 0 & 0\\ 0 & 0 & 2 & 0 & 0 & 2 & 0 & 0 & 0 & 0 & 2 & -1 & 2 & 0 & 0 & 2 & 0 & 0 & 0 & 0\\ 0 & 0 & 0 & 2 & 0 & 0 & 2 & 0 & 0 & 0 & 2 & 2 & -1 & 0 & 0 & 0 & 0 & 0 & 2 & 0\\ 0 & 0 & 2 & 0 & 0 & 0 & 0 & 2 & 0 & 0 & 0 & 0 & 0 & -1 & 2 & 2 & 2 & 0 & 0 & 0\\ 0 & 0 & 0 & 2 & 0 & 0 & 0 & 0 & 2 & 0 & 2 & 0 & 0 & 2 & -5 & 2 & 0 & 2 & 0 & 2\\ 0 & 0 & 0 & 0 & 0 & 0 & 0 & 0 & 0 & 2 & 0 & 2 & 0 & 2 & 2 & -1 & 0 & 0 & 2 & 0\\ 0 & 0 & 0 & 0 & 0 & 2 & 0 & 2 & 0 & 0 & 2 & 0 & 0 & 2 & 0 & 0 & -5 & 2 & 2 & 2\\ 0 & 0 & 0 & 0 & 0 & 0 & 2 & 0 & 2 & 0 & 0 & 0 & 0 & 0 & 2 & 0 & 2 & -1 & 2 & 0\\ 0 & 0 & 0 & 0 & 0 & 0 & 0 & 0 & 0 & 2 & 0 & 0 & 2 & 0 & 0 & 2 & 2 & 2 & -1 & 0\\ 0 & 0 & 0 & 0 & 0 & 0 & 0 & 0 & 0 & 2 & 0 & 0 & 0 & 0 & 2 & 0 & 2 & 0 & 0 & 3 \end{array}\right)
\label{Jmatrix}
\end{align}
\begin{align}
\mathcal{H}_{J'}=-J'\left(\begin{array}{cccccccccccccccccccc} 3 & 1 & 1 & 0 & 1 & 0 & 1 & 0 & 0 & 0 & 0 & 1 & 1 & 0 & 0 & 0 & 0 & 0 & 0 & 0\\ 1 & 1 & 0 & 0 & 0 & 0 & 0 & 0 & 1 & 0 & 0 & 0 & 0 & 1 & 1 & 0 & 0 & 0 & 0 & 0\\ 1 & 0 & 1 & 0 & 0 & 0 & 0 & 1 & 0 & 1 & 0 & 0 & 0 & 0 & 0 & 1 & 0 & 0 & 0 & 0\\ 0 & 0 & 0 &-1 & 0 & 0 & 0 & 0 & 1 & 0 & 0 & 0 & 0 & 0 & 0 & 1& 0 & 0 & 0 & 0\\ 1 & 0 & 0 & 0 & 1 & 0 & 0 & 1 & 0 & 0 & 0 & 0 & 0 & 0 & 0 & 0 & 1 & 1 & 0 & 0\\ 0 & 0 & 0 & 0 & 0 &-1& 0 & 0 & 0 & 0 & 0 & 0 & 0 & 0 & 0 & 0 & 0 & 0 & 1 & 0\\ 1 & 0 & 0 & 0 & 0 & 0 & 1 & 0 & 1 & 1 & 0 & 0 & 0 & 0 & 0 & 0 & 0 & 0 & 1 & 0\\ 0 & 0 & 1 & 0 & 1 & 1 & 0 & 1 & 0 & 0 & 0 & 0 & 0 & 0 & 0 & 0 & 0 & 0 & 0 & 1\\ 0 & 1 & 0 & 1 & 0 & 0 & 1 & 0 & 1 & 0 & 0 & 0 & 0 & 0 & 0 & 0 & 0 & 0 & 0 & 1\\ 0 & 0 & 1 & 0 & 0 & 0 & 1 & 0 & 0 &-1 & 0 & 0 & 0 & 0 & 0 & 0 & 0 & 0 & 0 & 0\\ 0 & 0 & 0 & 0 & 0 & 0 & 0 & 0 & 0 & 0 &-1 & 0 & 0 & 1 & 0 & 0 & 0 & 1 & 0 & 0\\ 1 & 0 & 0 & 0 & 0 & 0 & 0 & 0 & 0 & 0 & 0 & 1 & 0 & 1 & 0 & 0 & 1 & 0 & 1 & 0\\ 1 & 0 & 0 & 0 & 0 & 0 & 0 & 0 & 0 & 0 & 0 & 0 & 1 & 0 & 1 & 1 & 0 & 1 & 0 & 0\\ 0 & 1 & 0 & 0 & 0 & 0 & 0 & 0 & 0 & 0 & 1& 1 & 0 & 1 & 0 & 0 & 0 & 0 & 0 & 1\\ 0 & 1 & 0 & 0 & 0 & 0 & 0 & 0 & 0 & 0 & 0 & 0 & 1 & 0 &-1 & 0 & 0 & 0 & 0 & 0\\ 0 & 0 & 1 & 1 & 0 & 0 & 0 & 0 & 0 & 0 & 0 & 0 & 1 & 0 & 0 & 1 & 0 & 0 & 0 & 1\\ 0 & 0 & 0 & 0 & 1 & 0 & 0 & 0 & 0 & 0 & 0 & 1 & 0 & 0 & 0 & 0 &-1 & 0 & 0 & 0\\ 0 & 0 & 0 & 0 & 1 & 0 & 0 & 0 & 0 & 0 & 1 & 0 & 1 & 0 & 0 & 0 & 0 & 1 & 0 & 1\\ 0 & 0 & 0 & 0 & 0 & 1 & 1 & 0 & 0 & 0 & 0 & 1 & 0 & 0 & 0 & 0 & 0 & 0 & 1 & 1\\ 0 & 0 & 0 & 0 & 0 & 0 & 0 & 1 & 1 & 0 & 0 & 0 & 0 & 1 & 0 & 1 & 0 & 1 & 1 & 3 \end{array}\right)
\label{J'matrix}
\end{align}
\end{widetext}
\subsection{Zero spin-orbit coupling, $J'/J=0$.}
\label{sec:RVB1}
We now discuss the solution to the model. 
The ground state of the cluster obtained from the diagonalization of $\mathcal{H}$ in this basis:
\begin{equation}
|\Psi_0\rangle = \sum_m a_m |m \rangle,
\label{eq:psi0}
\end{equation}
where the coefficients (from larger to smaller weights) are: $-a_4=-a_6=a_{10}=-a_{11}=a_{15}=a_{17}=0.367328\equiv a$, $a_1=-a_{20}=0.144892\equiv b$, $a_2=a_3=a_5=a_7=-a_8=-a_9=a_{12}=a_{13}=-a_{14}=-a_{16}=-a_{18}=-a_{19}=0.111218\equiv c$. The symmetry in the coefficients can be attributed to the number of spin permutations $n_p$ needed in a triangle to get the other triangle spin configuration ($n_p=3$, $n_p=1$ and $n_p=3$ respectively). This is directly related with the number of connected states corresponding with the non-zero elements of \eqref{Jmatrix} + \eqref{J'matrix}. Moreover, it can be justified also from the symmetries presented in the system. While TRS, which switches both ${\bf k }\rightarrow {-\bf k }$ (IS) and $\sigma \rightarrow-\sigma$, is preserved in states with coefficients $a$ and $b$ (in this last one $C_3$ is kept too), for states with coefficients $c$ this invariance is lost. The ground state energy is $E^{Heis}_0=-5.30278 J$ ($J=4 t^2/U$) which matches very well with the exact solution of the Hubbard model for $U\gg t$: $E^{Hubb}(U=20t)=-1.031859t$  
($E^{Heis}_0=-1.060556t$), $E^{Hubb}(U=100t)=-0.211866t$ ($E^{Heis}_0=-0.212111t$).

We can compare our exact results in the $U\gg t$ limit with an RVB ansatz for the wavefunction\cite{Pauling,Anderson}.
We consider an RVB wavefunction which is a superposition of singlet configurations between 
nearest neighbor sites only. Hence, we only include the four configurations of Fig. \ref{QuantumSchemes} giving
the $| RVB \rangle$ state expressed in \eqref{eq:rvb}.
When expanding this state on the different VB configurations we get:
\begin{eqnarray}
\ket{RVB}&=&{ 1 \over \sqrt{17\times 4}} \left( |1 \rangle+|2 \rangle +|3 \rangle -3 |4 \rangle +  |5 \rangle  \right.
\nonumber \\
&-& 3 |6 \rangle +| 7 \rangle -| 8 \rangle -|9 \rangle + 3| 10 \rangle - 3 |11 \rangle 
\nonumber \\
&+& | 12 \rangle +| 13 \rangle - | 14 \rangle + 3 | 15 \rangle -| 16 \rangle  \nonumber\\
&+&  \left. 3 |17 \rangle| - | 18 \rangle - | 19 \rangle - | 20 \rangle\right).
\label{eq:developedrvb}
\end{eqnarray}
The overlap of the exact wavefunction with this $\ket{RVB}$ is $\langle RVB |\Psi_0 \rangle=0.9988$, indicating that $\ket{RVB}$ provides an accurate description of the Hubbard cluster in the $U\gg t$ limit. The energy associated with the $\ket{RVB}$ state: $\bra{RVB} \mathcal{H} \ket{RVB}=-5.29412 J$ which provides a very good estimate of the exact ground state energy only being a $0.16 \% $ higher. The tiny differences between the exact ground state and the $\ket{RVB}$ state can be attributed to neglecting singlets between next-nearest neighbors in $\ket{RVB}$ which favor the $S=3/2$ AF-type of configurations $\ket{1}$ and $\ket{20}$. 
The RVB describes the large anisotropy between the $n.n.$ spin correlations: $\langle RVB| {\bf S}_i {\bf S}_j |RVB \rangle ^{\bigtriangleup\rightarrow\bigtriangleup}=-{39 \over 17 \times 4}\approx -0.573$ and 
$\langle RVB | {\bf S}_i {\bf S}_j | RVB \rangle^{\bigtriangleup}  =-{13 \over  17 \times 4} \approx -0.191$ and  consistent with the ED calculations up to $N_s=42$ sites. The n.n.n. spin correlations are: $\langle RVB | {\bf S}_i {\bf S}_j | RVB \rangle= {36 \over 17 \times 16} \approx +0.132$, {\it i. e.} FM and
close to ($\approx +0.107$) in the $N_s=42$ site cluster.
\subsection{Finite spin-orbit coupling, $J'/J \neq 0$}
\label{sec:RVB2}
\begin{figure}[b!]
		\centering
		\begin{center}
			\includegraphics[width=6.cm]{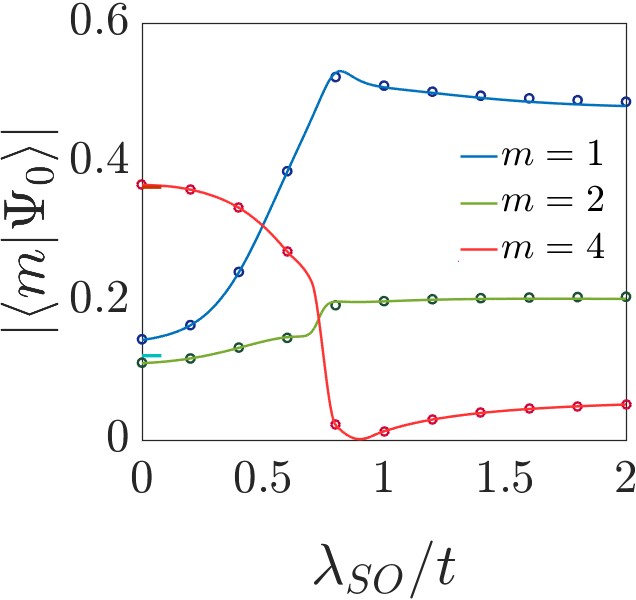}
		\end{center}
		\caption{Dependence of the ground state wavefunction on SOC at strong coupling, $U\gg t$. The modulus of the dominant non-ionic configurations of the wavefunction, $\abs{\langle {m}|\Psi_0 \rangle}$, contributing to the ground state of the Hubbard model with SOC on a six-site cluster for $U=100t$ are shown. A transition from the RVB state at $\lambda_{SO}\rightarrow 0$ to a $3/2$-MI like state occurs around $\lambda_{SO} \approx 0.8t$. The results from the Hubbard model (solid lines) are compared with the Heisenberg model (empty circles) showing very good agreement. We also show the coefficients from RVB at $\lambda_{SO}= 0$: $\abs{\langle {1}|RVB \rangle}=\abs{\langle {2}|RVB \rangle}$ and $\abs{\langle {3}|RVB \rangle}$ as cian and dark red dashes respectively \eqref{eq:developedrvb}.
		} 
 \label{HubbardHeisenberg}		
\end{figure}

In Fig. \ref{HubbardHeisenberg} we show the dependence of the main components of 
the ground state of the six-site cluster with spin-orbit coupling, $\lambda_{SO}$. The ground state of the Hubbard model for $U\gg t$ is compared with the ground state of the Heisenberg model (\ref{hjjp}) showing a good agreement, as it should. 
A transition to a state in which the configurations $\ket{1},\ket{20}$ dominate occurs around $\lambda_{SO}\sim 0.8t$, in good agreement with our results for $U=8t$ in Fig. \ref{coefficients}. These configurations are consistent with the $3/2$-MI state found in Hartree-Fock. However, in contrast to Hartree-Fock calculations, we find  
other configurations with non-negligible weight in the ground state. These are
associated with quantum fluctuations which effectively decrease the magnetic order of the pure classical $3/2$-MI state. Hence, our exact treatment of the model 
is consistent with the $3/2$-MI state but with a somewhat smaller staggered magnetic moment, {\it i. e.} $S \lesssim 3/2$. More careful work on larger clusters is
needed to confirm our prediction.
Note that due to the small size of the six-site cluster with PBC analyzed, it is necessary to take: $J'=4 (2\lambda_{SO})^2/U$ instead of $J'=4 \lambda_{SO}^2/U$ in evaluating the
Heisenberg model. This is because each lattice site is connected to a n.n.n. site by {\it two} hoppings of magnitude $\pm i \lambda_{SO}$ due to the PBC. This is equivalent to having the two n.n.n. sites connected by a {\it single} hopping, $\pm i 2\lambda_{SO}$, which is twice the original.


\begin{thebibliography}{99}
\bibitem{Discovered} M. Z. Hasan, and C. L. Kane, Rev. Mod. Phys., {\bf 82}, 3045 (2010).
\bibitem{KaneMele}  C. L. Kane, and E. J. Mele, Phys. Rev. Lett. {\bf 95}, 146802 (2005).
\bibitem{balents} W. Witczak-Krempa, G. Chen, Y. B. Kim, and L. Balents, Ann. Rev. Cond. Mat. Phys. {\bf 5}, 57 (2014).
\bibitem{pesin} D. Pesin, and L. Balents, Nat. Phys. {\bf 6}, 376  (2010).
\bibitem{kim} W. Witczak-Krempa, and Y. B. Kim, Phys. Rev. B {\bf 85}, 045124 (2012).
\bibitem{wen} X. G. Wen, {\it Quantum field theory of many-body systems: from the origin of sound to an origin of light and electrons}, Oxford University Press (2004).
\bibitem{lehur} S. Rachel, and K. Le Hur, Phys. Rev. {\bf 82}, 075106 (2010).
\bibitem{hohenadler} M. Hohenadler, and F. F. Assaad, J. Phys.: Condens. Matter {\bf 25} 143201 (2013).
\bibitem{assaad} M. Hohenadler, T. C. Lang, and F. F. Assaad,
Phys. Rev. Lett. 106, 100403 (2011).
\bibitem{merino} J. Merino, A. C. Jacko, A. L. Khosla, and B. J. Powell, Phys.
Rev. B {\bf 96}, 205118 (2017).
\bibitem{khosla} A. L. Khosla, A. C. Jacko, J. Merino, and B. J. Powell, Phys.
Rev. B {\bf 95}, 115109 (2017).
\bibitem{jacko}A. C. Jacko, A. L. Khosla, J. Merino, and B. J. Powell, Phys.
Rev. B {\bf 95}, 155120 (2017).
\bibitem{powell}B. J. Powell, J. Merino, A. L. Khosla, and A. C. Jacko, Phys.
Rev. B {\bf 95}, 094432 (2017).
\bibitem{iron3} Y.-Z. Zheng, M.-L. Tong, W. Xue, W.-X. Zhang, X.-M. Chen, F. Grandjean, and G. J. Long, Angew. Chem. Int. Ed. {\bf 46}, 6076 (2007).
\bibitem{henline} K. M. Henline, C. Wang, R. D. Pike, J. C. Ahern, B. Sousa, H. H. Patterson, A. T. Kerr, and C. L. Cahill, Crystal Growth \& Design {\bf 14}, 1449 (2014).
\bibitem{henling} L. M. Henling, and R. E. Marsh, Acta Crystallographica Section C {\bf 70}, 834 (2014).
\bibitem{cdmft} H.-F. Lin, Y.-H. Chen, H.-D. Liu, H.-S. Tiao, and W.-M. Liu, Phys. Rev. A {\bf 90}, 053627 (2014).
\bibitem{jacko1} A. C. Jacko, C. Janani, K. Koepernik, and B. J. Powell, Phys.
Rev. B {\bf 91}, 125140 (2015).
\bibitem{ruegg} A. R\"uegg, J. Wen, and G. A. Fiete, Phys. Rev. B {\bf 81}, 205115 (2010).
\bibitem{int} J. Wen., A. R\"uegg, C.-C. J. Wang, and G. A. Fiete, Phys. Rev. B {\bf 82}, 075125 (2010).
\bibitem{scarola} M. Chen, H.-Y.Hui, S. Tewari, and V. W. Scarola, Phys. Rev. B {\bf 97}, 035114 (2018).
\bibitem{manuel} M. F. L\'opez, and J. Merino, Phys. Rev. B {\bf 100} 075154, (2019).
\bibitem{Nourse} H. L. Nourse, R. H. McKenzie and B. J. Powell, arXiv:2003.04682v1 [cond-mat.str-el] (2020).
\bibitem{Heisenberg} J. Richter, J. Schulenburg, A. Honecker, and D. Schmalfuss, Phys. Rev. B {\bf 70}, 174454 (2004).
\bibitem{Orus} S. S. Jahromi, and R. Or\'us, Phys. Rev. B {\bf 98}, 155108 (2018).
\bibitem{kivelson} H. Yao and S. A. Kivelson, Phys. Rev. Lett. {\bf 99}, 247203 (2007).
\bibitem{Z2} L. Fu, and C. L. Kane, Phys. Rev. B. {\bf 76}, 045302 (2007).
\bibitem{Lieb} A. Bhattacharya, and B. Pal, Phys. Rev. B. {\bf 100} 235145 (2019) 
\bibitem{Fukuki} T. Fukuki, Y. Hatsugai, and H. Suzuki, J. Phys. Soc. Jpn. {\bf 74} (2005).

\bibitem{Haldane} F.D.M Haldane, Phys. Rev. Lett. {\bf 93}, 206602  (2004).
\bibitem{anderson87} P. W. Anderson, Science {\bf 235}, 1196 (1987).
\bibitem{Pauling} L. Pauling, and E.B. Wilson, Jr., {\it Introduction to Quantum Mechanics with Applications to Chemistry}, Dover, New York (1985)
\bibitem{Anderson} P. W. Anderson, Mater. Red. Bull. {\bf 8}, 153 (1973).













\end{thebibliography}
\end{document}